\documentclass[fleqn,usenatbib]{mnras}

\usepackage{amsmath}
\usepackage{amssymb}
\usepackage{color}
\usepackage{epic}
\usepackage{graphicx}
\usepackage{hyperref}
\usepackage{multirow}
\usepackage{natbib}
\usepackage{overpic}
\usepackage{orcidlink}
\usepackage{array}

\usepackage{color}
\AtBeginShipout{%
 \ifnum\value{page}>1 %
 \typeout{* Additional boxing of page `\thepage'}%
 \setbox\AtBeginShipoutBox=\hbox{\copy\AtBeginShipoutBox}%
 \fi
}

\title[RAiSE: {simulation}-based analytical model of AGNs]{RAiSE: {simulation}-based analytical model of AGN\\jets and lobes}

\author[R. J. Turner et al.]{Ross J. Turner$^{\,\orcidlink{0000-0002-4376-5455} \,1}$\thanks{Email: turner.rj@icloud.com}, Patrick M. Yates-Jones$^{\,\orcidlink{0000-0003-2806-3495}\,1}$, Stanislav S. Shabala$^{\,\orcidlink{0000-0001-5064-0493}\,1,2}$, Benjamin Quici$^{\,\orcidlink{0000-0002-7692-4934}\,3}$ and \and Georgia S. C. Stewart$^{\,\orcidlink{0000-0002-7155-6896}\,1}$\\
$^{1}$School of Natural Sciences, University of Tasmania, Private Bag 37, Hobart, 7001, Australia\\
$^{2}$ARC Centre of Excellence for All-Sky Astrophysics in 3 Dimensions (ASTRO 3D)\\
$^{3}$International Centre for Radio Astronomy Research, Curtin University, Bentley, WA 6102, Australia}

\date{Accepted 2022 October 17. Received 2022 October 11; in original form 2022 June 19}

\pubyear{2022}

\begin{document}

\label{firstpage}
\pagerange{\pageref{firstpage}--\pageref{lastpage}}
\maketitle

\begin{abstract}

We present an analytical model for the evolution of extended active galactic nuclei (AGNs) throughout their full lifecycle, including the initial jet expansion, lobe formation, and eventual remnant phases. A particular focus of our contribution is on the early jet expansion phase, which is traditionally not well captured in analytical models. We implement this model within the \emph{Radio AGN in Semi-Analytic Environments} (RAiSE) framework, and find that the predicted radio source dynamics are in good agreement with hydrodynamic simulations of both {low-powered} Fanaroff-Riley Type-I and {high-powered Type-II} radio lobes.
We construct synthetic {synchrotron surface brightness} images by complementing the {original} RAiSE model with the magnetic field and shock-acceleration histories of a set of Lagrangian tracer particles taken from an existing hydrodynamic simulation. We show that a single set of particles is sufficient for an accurate description of the dynamics and observable features of Fanaroff-Riley Type-II radio lobes with very different jet {parameters and ambient density profile normalisations}. Our new model predicts that the lobes of young ($\lesssim 10$\,Myr) sources will be both longer and brighter than expected at the same age from existing analytical models which lack a jet-dominated expansion phase; this finding has important implications for interpretation of radio galaxy observations.
{The RAiSE} code, written in {Python}, is publicly available on GitHub and PyPI.

\end{abstract}

\begin{keywords}
galaxies: active -- galaxies: jets -- radio continuum: galaxies
\end{keywords}

\section{INTRODUCTION}
\label{sec:INTRODUCTION}

The past two decades have unequivocally established the key role played by active galactic nucleus (AGN) jets in the cosmic evolution of galaxies and circumgalactic gas. On large scales, mechanical feedback by jet-inflated lobes suppresses runaway cooling in galaxy clusters \citep{Cowie+1977,Boehringer+1993,Fabian+2003,Forman+2005,Mittal+2009}, and maintenance-mode feedback is a standard feature of all modern galaxy formation models \citep[e.g.][]{Bower+2006,Croton+2006,Shabala+2009b,Fanidakis+2011,Dubois+2012,Schaye+2015,Raouf+2017,Weinberger+2018,Thomas+2021}. On galactic scales, observational and theoretical evidence exists for both negative \citep[i.e. suppression of star formation;][]{Nesvadba+2008,Dasyra+2012,Morganti+2013,Mukherjee+2021} and positive \citep[i.e. promotion of star formation;][]{Croft+2006,Tortora+2009,Crockett+2012,Dugan+2017} feedback.

Since their discovery in the 1960s, millions of AGN radio jets have been observed and catalogued; these numbers are set to increase dramatically in the near future thanks to \emph{Square Kilometre Array} pathfinder surveys such as ASKAP EMU \citep{Norris+2011}, LOFAR LoTSS \citep{Shimwell+2017,Shimwell+2019}, MeerKAT MIGHTEE \citep{Jarvis+2012} and MWA GLEAM \citep{Wayth+2015}. There are clear connections between the properties of emerging jets and their environments: the jets are predominantly found in rapidly cooling systems \citep{Best+2005,Rafferty+2006,Mittal+2009}, and their overall kinetic energy output appears to closely balance radiative cooling of the surrounding hot atmospheres. Yet many important questions remain unanswered. Jet activity is intermittent \citep{Schoenmakers+2000,Best+2005,Sabater+2019,Jurlin+2020,Quici+2021}, but the mechanisms responsible for this modulation are unknown; plausible candidates include magnetically arrested accretion \citep{Bisnovatyi-Kogan+1974,Tchekhovskoy+2011,Tchekhovskoy+2012}, chaotic cold accretion \citep{Gaspari+2017,McKinley+2022}, and jet-mediated feedback \citep{Bourne+2021,Husko+2022}. Similarly, while some studies \citep[e.g.][]{HK+2013,Bourne+2019} have begun addressing the question of how exactly the jets couple their energy to the circumgalactic gas, no detailed exploration exists of the relevant jet and environment parameter space.

Numerous lines of evidence point to a close relationship between the properties of jets and their environments. Core-brightened Fanaroff-Riley Type-I \citep[FR-I;][]{FR+1974} sources are found almost exclusively in dense environments; meanwhile, edge-brightened FR-IIs are more likely to be hosted by lower-mass galaxies \citep{Best+2012}, more commonly found in poor environments. Low-power FR-Is appear to be slowed down from initially relativistic speeds on galactic scales \citep{Laing+2012}, due to direct entrainment of interstellar gas \citep{Bicknell+1995} and/or stellar winds \citep{Perucho+2014,Wykes+2015}. Both FR-I and FR-II jets are capable of inflating large `classical double' radio lobes, however {X-ray constraints on the pressure of the surrounding ambient medium} show conclusively that, unlike in FR-IIs, the FR-I lobes must contain a substantial fraction of non-radiating particles {to enable pressure balance} \citep{Croston+2014,Ineson+2017}. Although observed radio sources exhibit a wide variety of apparent structures, these can be readily explained by a combination of large-scale dynamics \citep[e.g.][]{Hardcastle+2019b,Missaglia+2019} and projection effects \citep[e.g.][]{Harwood+2020}. Analytical radio source models therefore provide an important tool for understanding the physics of these objects.

Environment-sensitive models of classical double radio sources date back to the `twin-exhaust' model of \citet{Blandford+1974} and the `dentist drill' model of \citet{Scheuer+1974}, and have been subsequently extended by several authors \citep[e.g.][]{Begelman+1989,Falle+1991,KA+1997,BR+2000,Manolakou+2002,Turner+2015,Hardcastle+2018}. In these models, first applied to FR-II sources (but see below), the initially ballistic jets are collimated first by the surrounding gas and later by the lobe (or cocoon) formed due to backflow of jet material from the jet termination shock. The resultant lobe expands supersonically, driving a bow shock through the surrounding gas. Slow, low-power jets may run out of forward thrust before recollimation, producing FR-I sources either with or without lobes \citep{Alexander+2006,Krause+2012}. The dynamics of lobed FR-Is (i.e. those FR-Is which have formed a lobe) are globally similar to FR-IIs: both source types can drive strong bow shocks into the surrounding gas due to the large pressure mismatch between the overpressured lobe and ambient medium. The main difference is in the spatial distribution of synchrotron-emitting particles \citep{Turner+2018a}: in FR-Is, shock-accelerated particles flow forward from the flare point, while in FR-IIs the lobe is inflated by backflow from the hotspots, resulting in different spectral signatures. \citet{KA+1997} presented the first radio source model in non-constant density environments, and showed that for power-law atmospheres (i.e. external pressure $p_x(r) \propto r^{-\beta}$ for some constant $\beta$), the lobe expansion will be self-similar -- in other words, the lobe will maintain a fixed length-to-width ratio. Extending this analysis, \citet{Turner+2015} showed that this is no longer the case in more complex environments, with older radio sources expected to be more elongated, a result consistent with observations. 

While the analytical models described above provide an excellent description of late-time lobe evolution, hydrodynamic and magnetohydrodynamic simulations \citep{Sutherland+2007,Krause+2012,HK+2013,Yates+2021} highlight another important, earlier phase of radio source evolution. This `jet breakout' phase \citep{Sutherland+2007} corresponds approximately to the {propagation of the jets into the circumgalactic medium, before any recollimation and subsequent lobe formation on larger scales \citep{Krause+2012}}. While analytical models exist for this jet-dominated expansion phase \citep{Alexander+2006}, at present no analytical radio source model adequately describes \emph{both} the early jet phase, and subsequent supersonic lobe expansion. We aim to address this issue in the present work.

Our starting point is the \emph{Radio AGN in Semi-Analytic Environments} (RAiSE) model, which successfully models the evolution of jet-inflated Fanaroff-Riley Type-I and -II lobes in complex environments. In previous work, RAiSE was used to quantify the jet energy budget in low-redshift AGNs \citep{Turner+2015, Turner+2018b}, address the discrepancy between spectral and dynamical radio source ages \citep{Turner+2018a}, probe the duty cycle of radio galaxies \citep{Turner+2018,Shabala+2020,Quici+2022}, make predictions for future X-ray surveys \citep{Turner+2020a}, search for distant radio galaxies \citep{Turner+2020b}, and test cosmological models \citep{Turner+2019}. In the present contribution, we first develop the formalism to describe early, jet-dominated expansion phases of radio source evolution, and the transition to the standard lobe phase (Section~\ref{sec:ANALYTIC DYNAMICAL MODEL}); we incorporate these into the existing RAiSE model which already considers the late-time evolution of active and remnant sources. In Section~\ref{sec:Particle emissivity model}, we present our method to create synthetic surface brightness images using the magnetic field and shock-acceleration histories of a set of Lagrangian tracer particles taken from an existing hydrodynamic simulation; these particles are adapted to the dynamics of the analytical model.
We compare model predictions with detailed hydrodynamic simulations in Section~\ref{sec:ANALYSIS OF DYNAMICAL MODEL PREDICTIONS}, and discuss the implications of our results in Section~\ref{sec:Jet contribution to evolutionary history}. We conclude in Section~\ref{sec:CONCLUSIONS}.

Throughout the paper, we assume a $\Lambda \rm CDM$ concordance cosmology with $\Omega_{\rm M} = 0.3$, $\Omega_\Lambda = 0.7$ and $H_0 = 70 \rm\,km \,s^{-1} \,Mpc^{-1}$ \citep{Planck+2016}.

\section{ANALYTIC DYNAMICAL MODEL}
\label{sec:ANALYTIC DYNAMICAL MODEL}

\begin{figure*}
\begin{center}
\includegraphics[width=0.6\textwidth,trim={0 0 0 0},clip]{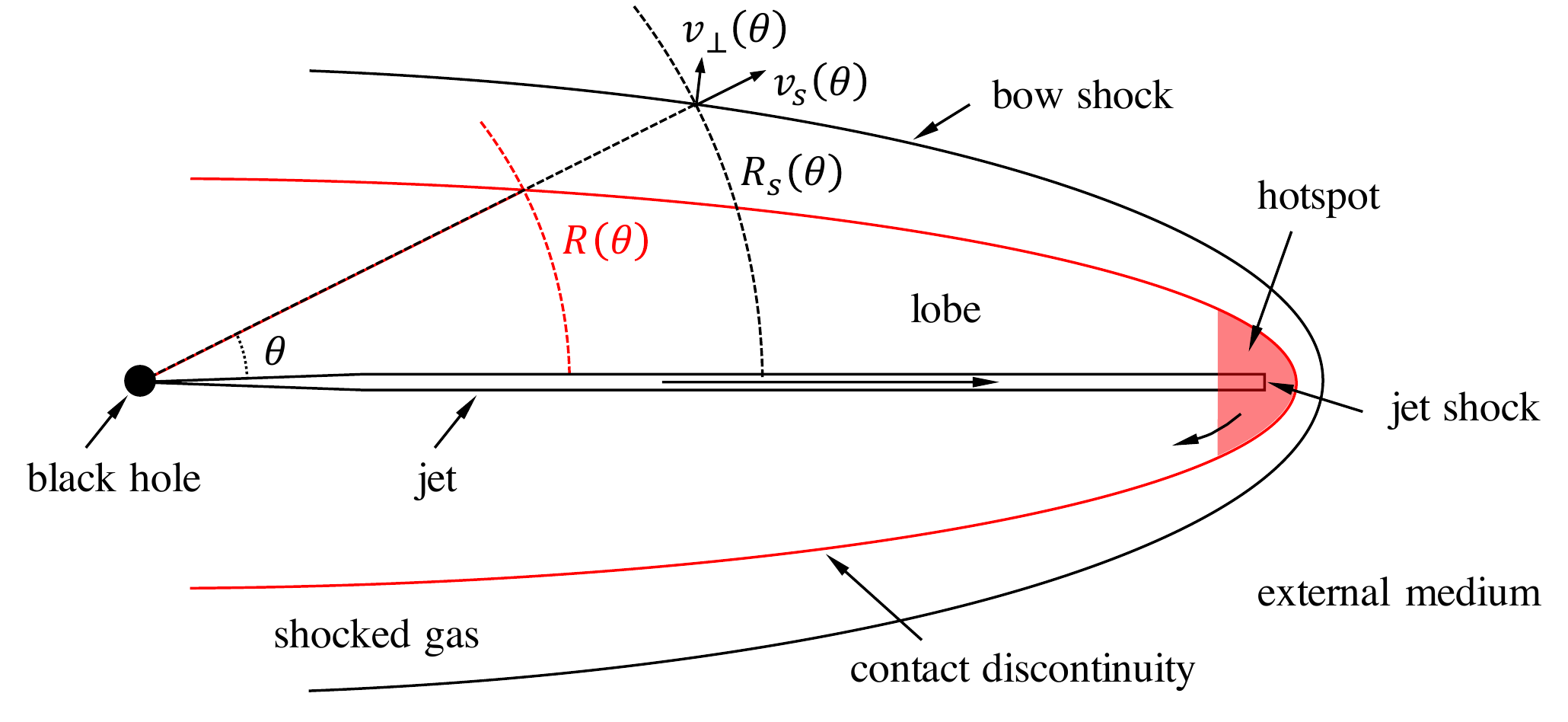}
\end{center}
\caption[]{Schematic of the dynamical model for the lobe and shocked shell.}
\label{fig:dynamics}
\end{figure*}

The analytical model used in {this work} is based on the modified version of the {original} {RAiSE} dynamical model, as described by \citet{Turner+2020a}. Their dynamical model assumes a high-powered relativistic plasma jet drills through the ambient medium generating a bow shock which radiates outwards from the jet-head (Figure \ref{fig:dynamics}; shown for the jet on one side of the active nucleus). The bow shock overruns the ambient medium greatly increasing its pressure as described by the Rankine–Hugoniot jump conditions for a plane-parallel shock. Meanwhile, the plasma in the jet is shock-accelerated at the jet-head (first-order Fermi acceleration) and is forced backwards towards the active nucleus by the pressure of the shocked ambient gas inside the bow shock; the region filled by the shock-accelerated plasma is referred to as the lobe. 

In this section, we detail the necessary changes to the dynamical model to consider the initial jet breakout phase (Section \ref{sec:Relativistic plasma jet}), the subsequent inflation of the lobe (Section \ref{sec:Two phase fluid}), and the late-time evolution of the lobe and surrounding shocked gas shell (Section \ref{sec:Lobe and shocked gas shell}). We begin by summarising the parameterisation of a general ambient medium in the RAiSE framework (Section \ref{sec:Ambient density and temperature profiles}).

\subsection{Ambient density and temperature profiles}
\label{sec:Ambient density and temperature profiles}

The {RAiSE} models assume spherically symmetric external environments for each lobe. The gas density and temperature profiles are approximated by $n$ power-laws spaced uniformly in log-radius from 1\% to 100\% of the cluster virial radius. The power-law approximations for the density and temperature are given by,
\begin{equation}
\begin{split}
\rho_x(r) &= k_ir^{-\beta_i}  \\
\tau_x(r) &= l_i \frac{\bar{m}}{k_b} r^{-\xi_i} 
\end{split} \ \ \ , \ r_{i} \leqslant r \leqslant r_{i+1} \ \ \ , 0 \leqslant i < n
\label{profile}
\end{equation}
where the exponents $\beta_i$ and $\xi_i$ are set by fitting power-laws to the values of the complete density and temperature profiles at $r_i$ and $r_{i + 1}$. The constants $k_i$ and $l_i$ are set to ensure continuity of the $i$-th approximating power-law with the preceding power-law. The value of these constants in the first power-law are $k_0 = \rho_0 a^{\beta_0}$ and $l_0 = \tau_0 \frac{k_b}{\bar{m}} a^{\xi_0}$ for some radius $a$ with known density and temperature $\rho_0$ and $\tau_0$ respectively. Here, $k_b$ is the Boltzmann constant and $\bar{m} \sim 0.6m_p$ is the average particle mass of the plasma comprising the ambient medium, for proton mass $m_p$.

The temperature profile is generally well approximated by a constant value, at least to within a factor of 2-3 for clusters \citep{Vikhlinin+2006}; to reduce complexity in adding hydrodynamic simulation particles to the analytical model we hereafter assume $\xi_i = 0$ for all $i$.

\subsection{Relativistic plasma jet}
\label{sec:Relativistic plasma jet}

The bi-polar relativistic plasma jets emanating from the active nucleus accelerate leptonic particles to near the speed of light ($\gamma \gg 1$) through their strong helical magnetic fields. The passage of the jet outwards from the core is greatly impeded by the multiphase medium of the host galaxy, with hydrodynamic simulations \citep{Sutherland+2007, Mukherjee+2016} showing a travel time of order 0.1-1\,Myr to escape the galactic bulge. However, upon clearing a path through the multiphase medium, the forward expansion of the jet through the cluster becomes comparable to the relativistic speeds of the bulk plasma flow. We model the expansion of the jet upon escaping its host galaxy as follows.

\subsubsection{Relativistic hydrodynamic equations}
\label{sec:Relativistic hydrodynamic equations}

The relativistic hydrodynamic equations can relate the properties of fluids upstream and downstream of a shock discontinuity in terms of the stress-energy tensor. However, these equations can be greatly simplified for a relativistic fluid with the shock front expanding along a single spatial dimension.
The conservation equations for a relativistic fluid along this single spatial dimension are expressed in terms of comoving quantities such as gas density $\rho$, gas pressure $p$, specific enthalpy $hc^2$ (i.e. $h$ is dimensionless), and the non-zero spatial component of the four-velocity $u = \gamma v/c$ (hereafter shortened to four-velocity) relative to the shock front. 

The conservation equations for a relativistic fluid are as follows \citep[e.g.][]{Walg+2013, Fukue+2018}:
\begin{subequations}
\begin{gather}
\rho \gamma v = \rho_1 \gamma_1 v_1 \ \ \ \text{(continuity)} \label{continuity}\\
\rho h \gamma^2 v^2 + p = \rho_1 h_1 \gamma_1^2 v_1^2 + p_1 \ \ \ \text{(momentum)} \label{momentum} \\
\rho (h \gamma - 1) \gamma v = \rho_1 (h_1 \gamma_1 - 1) \gamma_1 v_1  \ \ \ \text{(energy)}.
\end{gather}
\end{subequations}
The fluid downstream of the shock is represented by the subscript `1' whilst no subscript refers to the upstream fluid.

For a polytropic equation of state (EoS) in the fluid, the dimensionless specific enthalpy is \citep{Mignone+2007}:
\begin{equation}
h = 1 +  \frac{\Gamma}{\Gamma - 1}\frac{p}{\rho c^2} ,
\label{enthralpy}
\end{equation}
where $\Gamma$ is the polytropic index (or adiabatic index for an adiabatic EoS). In this work, we assume the jet plasma is not relativistically hot; i.e. $h \approx 1$ for kinetically dominant jets.

We derive a Rankine-Hugoniot relation for the density and velocity of the jet plasma (upstream fluid) and ambient medium of the jets' environment (downstream fluid) using conservation of momentum flux (Equation \ref{momentum}). 
The bulk velocity of the ambient medium in the observer frame is zero at all times for random particle motions. As a result, the bulk velocity of these particles in the frame of the shock front, $v_1$, is exactly equal to the expansion rate of the shock in the observer frame, $v_s$ (see Figure \ref{fig:dynamics}); i.e. $v_1 \equiv -v_s$, with corresponding Lorentz factor $\gamma_1 \equiv \gamma_s$. 
By contrast, the spatially-averaged (across the jet cross-section) bulk velocity of the upstream fluid particles in the jet is non-zero, defined as $\bar{v}_j$ in the observer frame with corresponding Lorentz factor $\bar{\gamma}_j$. The product of the Lorentz factor and velocity for the jet particles in the frame of the shock front is related to the rest- and observer-frame quantities as $\gamma v = \bar{\gamma}_j \gamma_s (\bar{v}_j - v_s)$.
The conservation of momentum flux equation can therefore be rewritten as:
\begin{equation}
\bar{\rho}_j h_j \bar{\gamma}_j^2 \gamma_s^2 (\bar{v}_j - v_s)^2 = \rho_x h_x \gamma_s^2 v_s^2 ,
\label{jethead}
\end{equation}
where $h_j$ is the dimensionless specific enthalpy of the jet, and $\rho_x$ and $h_x$ are the density and dimensionless specific enthalpy of the (external) ambient medium respectively. Here, we have assumed the jet and ambient medium are in approximate thermal pressure equilibrium, or at least any difference is negligible compared to magnitude of the ram pressure components. This conservation equation is rearranged to yield a relationship between the jet-head advance speed and the bulk velocity of the jet as \citep[cf.][]{Marti+1993}:
\begin{equation}
v_s = \frac{\bar{v}_j}{1 + [\bar{\rho}_j h_j \bar{\gamma}_j^2/(\rho_x h_x)]^{-1/2}} ,
\label{vs}
\end{equation}
where we define the dimensionless quantity $\eta_R = \bar{\rho}_j h_j \bar{\gamma}_j^2/(\rho_x h_x)$, which is a function of properties of the jet and ambient medium, including the jet kinetic power, as we now describe.

Following \citet{Walg+2013}, the kinetic power of the jet, $Q$, is related to its rest-mass energy discharge, $\dot{M}c^2$, and the spatially-averaged Lorentz factor of the flow, $\bar{\gamma}_j$,  at some location $r$ along the jet as $Q = (h_j \bar{\gamma}_j - 1) \dot{M} c^2$. 
Meanwhile, the rest-mass energy discharge is related to the spatially-averaged density of jet, $\bar{\rho}_j(r)$, its four-velocity, $\bar{u}_j = \bar{\gamma}_j \bar{v}_j$, and cross-sectional area, $\Omega r^2$, as $\dot{M} = \bar{\rho}_j(r) \bar{u}_j \Omega r^2$. 
The solid angle of the jet is $\Omega = 2\pi(1 - \cos\theta_j)$, where $\theta_j$ is the `apparent' half-opening angle of the jet; this apparent half-opening angle is defined as $\theta_j = \arctan(r_h/r)$ for jet length $r$ and jet radius at the hotspot $r_h$.
The spatially-averaged density of the jet is therefore defined as:
\begin{equation}
\bar{\rho}_j(r) = \frac{Q}{2\pi \bar{u}_j c^2 (h_j \bar{\gamma}_j - 1) (1 - \cos\theta_j) r^2} .
\label{jetdensity}
\end{equation}
The ratio of the jet and ambient densities in Equation \ref{vs} for the jet-head advance speed can thus be written as follows:
\begin{equation}
\eta_R(r) = \frac{Q h_j \bar{\gamma}_j^2}{2 \pi k h_x \bar{u}_j c^2 (h_j \bar{\gamma}_j - 1)(1 - \cos\theta_j) r^{2 - \beta}} ,
\label{L}
\end{equation}
where we have used the expression for the density of the ambient medium in Equation \ref{profile}. Setting $\eta_R(r) = 1$ gives the length-scale at which the jet density first falls below that of the ambient medium. This metric is therefore equivalent to the $L_{1b}$ length-scale of \citet{Krause+2012}, but for a relativistic jet and general ambient medium.

\subsubsection{Jet spine and sheath}
\label{sec:Jet spine and sheath}

\begin{figure*}
\begin{center}
\includegraphics[width=0.65\textwidth,trim={0 0 0 0},clip]{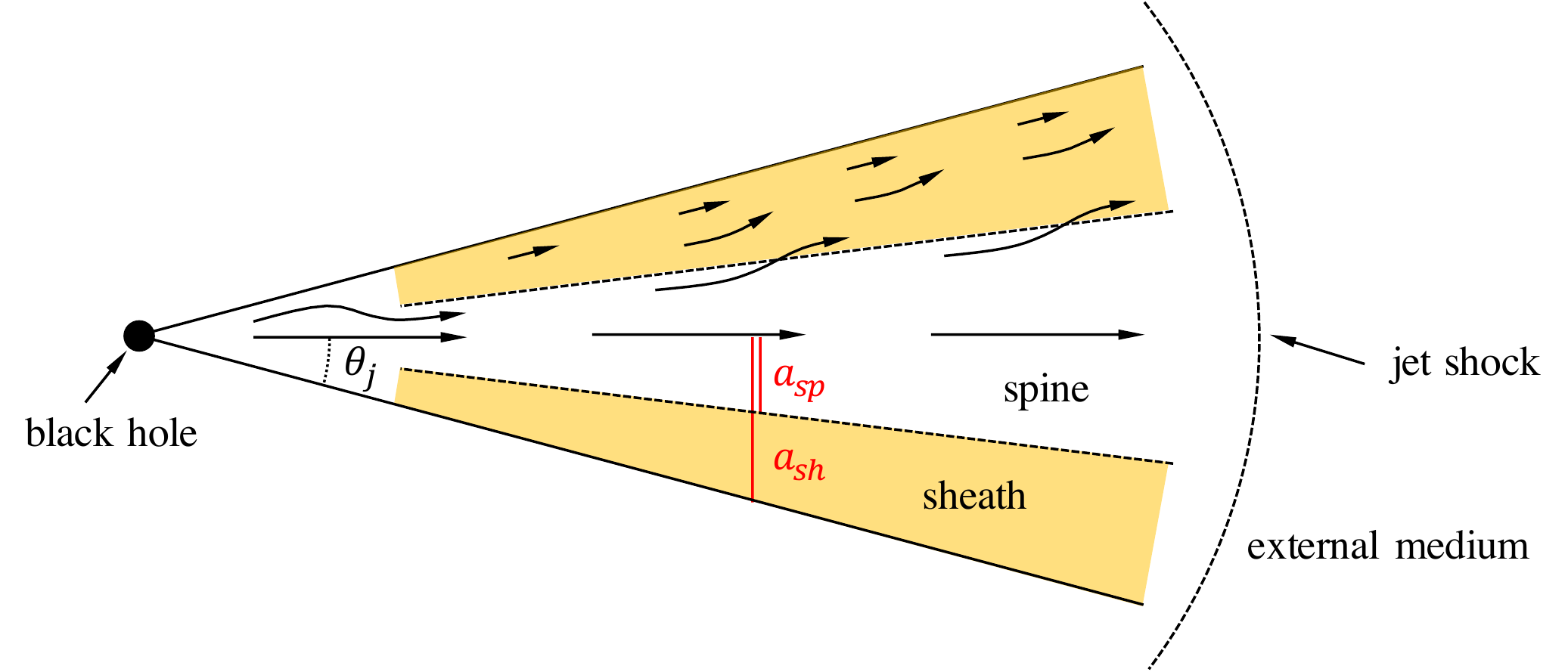}
\end{center}
\caption[]{Schematic of the dynamical model for the initially un-collimated relativistic jet.}
\label{fig:jet_dynamics}
\end{figure*}

The relativistic jet is modelled as a conical structure with a high-velocity (along the jet) and low-density spine, surrounded by a low-velocity and high-density sheath (Figure \ref{fig:jet_dynamics}). The bulk velocity of plasma injected into the jet is assumed to be spatially uniform \citep[as in hydrodynamic simulations; e.g.][]{Yates+2018}, but slows near the jet edge due to the formation of a boundary layer. The jet spine in M87 is observed to have bulk velocities, $u$, at least three times greater than the jet sheath at 100-1000 Schwarzchild radii \citep{Mertens+2016}, with this ratio expected to increase at larger distance scales. The continuity equation (Equation \ref{continuity}) states that in the absence of sidewards motions the density in this low-velocity region must increase markedly leading to an accumulation of mass along the jet edge. This conclusion is consistent with arguments based on deriving the cross-sectional pressure profile of the jet; for example, \citet{Walg+2013} finds that an isochoric jet (similar, but non-linear, finding for isothermal jet) has a step function in the plasma density with a modest factor of approximately five between the spine and sheath. 

The circular cross-section of the jet is therefore, in this work, assumed to comprise a spine of constant density $\rho_{s\!\;\!p}(r)$ within cross-sectional radius $a_{s\!\;\!p}$. Similarly, the sheath has constant density $\rho_{s\!\;\!h}(r)$ between the spine and outer edge of the jet at cross-sectional radius $a_{s\!\;\!h}$ (see Figure \ref{fig:jet_dynamics}). The conical expansion of the jet causes both the sheath and spine densities to reduce with increasing jet length as $1/r^2$. By contrast, the four-velocities of the bulk fluid flow in the spine and sheath are unaffected by this expansion according to the conservation equations. The commencement of lobe formation, at a typical distance scale of order 1-10\,kpc, is shortly followed by the collimation of the jet due to the increase in thermal pressure (from the lobe plasma) directly surrounding the side of the jet \citep[e.g.][]{Krause+2012}. The cross-sectional radius of the jet will increase only modestly whilst collimated, however, the continuing lobe formation will rapidly transition the expansion from jet-dominated to lobe-dominated. As a result, we parameterise the two co-dependent events, the onset of lobe formation and jet collimation, through a single variable: the apparent half-opening angle of the jet at the start of the lobe-dominated expansion phase (see Section \ref{sec:Two phase fluid} for detailed discussion). Specifically, the conical jet is assumed to have a constant half-opening angle, $\theta_j$, throughout the jet-dominated expansion phase, with this apparent opening angle calibrated based on the dynamics of hydrodynamic simulations.

We now relate the bulk velocities in the spine and sheath by deriving the spatially-averaged four-velocity across the jet cross-section. That is,
\begin{equation}
\begin{split}
\bar{u}_j &= \bigg[\, \int_0^{a_{s\!\;\!p}} u_{s\!\;\!p} a da + \int_{a_{s\!\;\!p}}^{a_{s\!\;\!h}} u_{s\!\;\!h} a da\, \bigg] \bigg /  \int_0^{a_{s\!\;\!h}} a da \\
&= u_{s\!\;\!p} \left(\frac{a_{s\!\;\!p}}{a_{s\!\;\!h}}\right)^2 + u_{s\!\;\!h} \bigg[1 - \left(\frac{a_{s\!\;\!p}}{a_{s\!\;\!h}}\right)^2 \bigg] ,
\end{split}
\end{equation}
where we assume the bulk velocity of the spine is much greater than that of the sheath, i.e. $u_{s\!\;\!p} \gg u_{s\!\;\!h}$. The spatially-averaged four-velocity of the jet therefore tends to $\bar{u}_j = a_{*}^2 u_{s\!\;\!p}$, where we define $a_{*} = a_{s\!\;\!p}/a_{s\!\;\!h} < 1$ for compactness  in later sections. The spatially-averaged bulk velocity and Lorentz factor of the flow are thus related to the Lorentz factor of the jet spine as follows:
\begin{subequations}
\begin{gather}
\bar{v}_j = \bigg[ \frac{\gamma_{s\!\;\!p}^2 a_{*}^4 - a_{*}^4}{\gamma_{s\!\;\!p}^2 a_{*}^4 - a_{*}^4 + 1} \bigg]^{1/2} c \label{jetvelocity}\\
\bar{\gamma}_j = \big[ \gamma_{s\!\;\!p}^2 a_{*}^4 - a_{*}^4 + 1 \big]^{1/2} .
\end{gather}
\end{subequations}
The bulk flow of the jet is parameterised in terms of the Lorentz factor of the spine in this work. In Section \ref{sec:ANALYSIS OF DYNAMICAL MODEL PREDICTIONS} onwards, we refer to this Lorentz factor as $\gamma_j \equiv \gamma_{s\!\;\!p}$ for consistency with the equivalent definition presented in hydrodynamic simulations.

\subsubsection{Relativistic differential equations}

The equation governing the expansion of the jet-head, $\dot{R}_s \equiv v_s$ (Equation \ref{vs}), is differentiated to produce an equation for the acceleration of the contact discontinuity.
Following \citet{Turner+2015} for their non-relativistic lobe solution, we adopt a numerical scheme using a fourth-order Runge-Kutta method in terms of a system of three coupled ODEs for the displacement, velocity and Lorentz factor (cf. Equation \ref{supersonic system}). This decision enables the same computational framework used in their work for the lobe-dominated expansion phase to be applied throughout the source evolutionary history of the source.
For angles within the half-opening angle of the jet ($0 \leqslant \theta \leqslant \theta_j$), we have
\begin{equation}
\begin{split}
\dot{R}_s(\theta) &= v_s \\
\dot{v}_s(\theta) &= \frac{(\beta - 2) \bar{v}_j v_s}{2 R_s \eta_R^{1/2} [1 + \eta_R^{-1/2}]^2} \\
\dot{\gamma}_s(\theta) &= \frac{\gamma_s^3 v_s \dot{v}_s}{c^2} ,
\end{split}
\label{jet system}
\end{equation}
where we assume the jet radius is otherwise independent of angle $\theta$ for the small opening angles of typical jets. The velocity of the jet-head is updated at each time step directly from the Lorentz factor at highly-relativistic speeds to avoid numerical errors.

\subsection{Lobe and shocked gas shell}
\label{sec:Lobe and shocked gas shell}

The shocked gas shell and lobe comprising synchrotron-emitting electrons are modelled following the work of \citet{Turner+2015} and \citet{Turner+2020a}. We summarise the equations used in their dynamical models and specify modifications to comply with the relativistic formulation used for the jet, as discussed in the previous section.

\subsubsection{Shocked gas shell}

The bow shock of our dynamical model is an ellipsoid with the major axis aligned along the axis of the jet and the two minor axes having equal length; i.e. the bow shock has a circular cross-section. The ratio of the lengths of the major and minor axes is defined as the axis ratio of the shocked shell, $A_s$; however, we generally will refer to the axis ratio of the lobe in this work as it is an observable quantity through synchrotron-emission. 

The shell of shocked gas is partitioned into an ensemble of small angular elements which are each assumed to receive a time-invariant fraction of the jet power as the bow shock expands. This assumption yields self-similar expansion when the bow shock is expanding supersonically (in a power-law gas density profile), albeit the shocked shell elongates slowly due to a steepening ambient medium. Following \citet{Turner+2015}, the volume of each small angular element of the shocked shell, $[\theta - \delta\theta/2, \theta + \delta\theta/2)$, is given by,
\begin{equation}
\delta V(\theta) = \frac{2\pi R_s^3(\theta)}{3} \sin \theta \delta\theta ,
\end{equation}
where $\theta$ is the angle between some location on the surface of the shocked shell and the jet axis, and $R_s(\theta)$ is the radius of the shell at that location (Figure \ref{fig:dynamics}).

For the assumed ellipsoidal geometry, the radial distance to the surface of the shocked shell at angle $\theta$ is related to the length of the shell along the jet axis, $R_s(\theta = 0)$, by the ratio given in Equation 10 of \citet{Turner+2020a},
\begin{equation}
\eta_s(\theta) = \frac{1}{\sqrt{A_s^{2} \sin^2 \theta + \cos^2 \theta}} .
\label{eta_s}
\end{equation}
The component of the expansion rate normal to the surface of the bow shock, $v_\perp(\theta)$, is related to the expansion rate of the shell along the jet axis, $v_\perp(\theta = 0) \equiv v_s(\theta = 0)$, by the ratio given in Equation 11 of \citet{Turner+2020a},
\begin{equation}
\zeta_s(\theta) = \left[\frac{A_s^{2} \sin^2 \theta + \cos^2 \theta}{A_s^{4} \sin^2 \theta + \cos^2 \theta} \right]^{1/2} .
\label{zeta_s}
\end{equation}
These equations are used to parameterise the Mach number of the bow shock at each location on the surface of the shocked shell in terms of the radius of the shell along a single axis (i.e. the jet axis), and permit an analytic solution; i.e. $M(\theta) = v_\perp(\theta)/c_x$, where $c_x$ is the sound speed of the ambient medium, constrained by the gas density and temperature profiles.

\subsubsection{Shock-accelerated plasma lobe}
\label{sec:Shock-accelerated plasma lobe}

Hydrodynamic simulations of powerful radio sources \citep[e.g.][]{HK+2013} suggest that, once the lobe is fully formed, the relative sizes of the lobe and bow shock remain constant (at least for strong-shock/supersonic expansion). Following \citet{Turner+2020a}, we therefore model the growth of the shocked shell and lobe through a time-invariant constant of proportionality, $b(\theta)$; this constant of proportionality may be a function of the angle between the surface location and the jet axis, $\theta$. Simulations clearly show that the ratio between the bow shock and lobe radii varies across the surface of the shocked shell: the lobe is typically closest to the bow shock near the hotspot. We define the axis ratio of the bow shock in terms of the observable axis ratio of the fully-formed radio lobe, $A$, as $A_s = A^\iota$, for some exponent $\iota$. In this work, this parameter $\iota$ is calibrated using hydrodynamic simulations.

Similar to the shocked shell, the radial distance to the lobe surface at an angle $\theta$ is related to the length of the lobe along the jet axis, $R(\theta = 0)$, as:
\begin{equation}
\eta(\theta) = \frac{1}{\sqrt{A^2 \sin^2 \theta + \cos^2 \theta}} .
\label{eta}
\end{equation}
The radius of the shocked shell can thus be related to that of the lobe as $R_s(\theta) = b \eta_{\rm s}(\theta) R(\theta)/\eta(\theta)$, where $b \equiv b(\theta=0)$ is the ratio of the shocked shell to lobe radii along the jet axis.

\subsubsection{Relativistic fluid dynamics}
\label{sec:Relativistic fluid dynamics}

The expansion of the shocked shell (and implicitly the lobe) is modelled by \citet{Turner+2020a} assuming the sound crossing time of the shell is much less than the dynamical age; i.e. the thermal pressure of the plasma predominantly drives the expansion. In this work, we similarly describe the expansion of the shocked shell using the following relationship for adiabatic expansion of a small angular volume element with momentum flux $p_s(\theta)$ and volume $\delta V_s(\theta)$ \citep[see Equation 6 of][]{Alexander+2006}:
\begin{equation}
\frac{dp_s(\theta)}{dt}\delta V_s(\theta) + \Gamma_c p_s(\theta) \frac{d[\delta V_s(\theta)]}{dt} = (\Gamma_c - 1)Q \delta \lambda(\theta) ,
\label{gov}
\end{equation}
where $\Gamma_c = \tfrac{5}{3}$ is the adiabatic index of the shocked gas and lobe plasma, $Q$ is the power injected into the shocked shell by the jet, and $\delta \lambda(\theta)$ is the fraction of that power associated with the expansion of the volume $\delta V_s(\theta)$.

The momentum flux at the bow shock (at angle $\theta$) is related to the properties of the ambient medium through the relativistic conservation of momentum equation (Equation \ref{momentum} as:
\begin{equation}
p_s(\theta) = \rho_x(\theta) h_x \gamma_s^2(\theta) v_s^2(\theta) + p_x(\theta) ,
\end{equation}
where the density and pressure of the ambient medium are defined through the gas density and temperature profiles, $\rho_x = kr^{-\beta}$ and $p_x = (kl) r^{-\beta}$, whilst the dimensionless specific enthalpy of the non-relativistic ambient medium is assumed to be $h_x \approx 1$. Importantly, the momentum flux in the shocked shell includes components due to both ram and thermal pressure; i.e. $p_s(\theta) = p_{r\!\;\!a\!\;\!m}(\theta) + p(\theta)$. These cannot be separated as for the jet model as the thermal pressure in the lobe, $p(\theta)$, is comparable to the forward ram pressure of the jet, $p_{r\!\;\!a\!\;\!m}(\theta)$, and will not completely dominate the momentum flux until the jet switches off.

Following \citet{Turner+2015}, we write the momentum flux of the shocked shell in terms of its radius at angle $\theta$, $R_s(\theta)$, and properties of the ambient medium as follows:
\begin{equation}
p_s(\theta) = k R_s^{-\beta}(\theta) [\zeta_s \gamma_s \dot{R}_s/ \eta_s]^2(\theta) + (k l) R_s^{-\beta}(\theta) ,
\label{plobe}
\end{equation}
where $\gamma_s(\theta)$ is the Lorentz factor of the shocked shell expanding with relativistic velocity $\dot{R}_s(\theta)$, and the derivatives are with respect to the time in the observer frame, $t$. This relation, derived from the relativistic conservation equations in Section \ref{sec:Relativistic hydrodynamic equations}, is consistent with the non-relativistic Rankine-Hugoniot jump conditions for a plane parallel shock.

The derivatives of the momentum flux and volume in the governing differential equation are now much more complicated than in the previous iterations of RAiSE due to the Lorentz factor terms. Here, in our presentation of the theory, we will only consider the strong-shock supersonic expansion limit where the relativistic equations differ considerably to those previously derived by \citet{Turner+2015}.
The first-order derivative of the pressure in the strong-shock supersonic limit is thus given by,
\begin{equation}
\begin{split}
\frac{dp_s(\theta)}{dt} &= k R_s^{-\beta}(\theta) \dot{R}_s(\theta) [\zeta_s \gamma_s/ \eta_s]^2(\theta)\\
&\quad\quad \times \left[ -\beta R_s^{-1} \dot{R}_s^2 + 2  \ddot{R}_s [1 + (\gamma_s \dot{R}_s/c)^2] \right]\!(\theta) .
\end{split}
\end{equation}

The differential equation governing the expansion of the shocked shell (Equation \ref{gov}) is rewritten in terms of the size, velocity and acceleration of the shell, in addition to properties of the ambient medium, for the relativistic fluid dynamics equations. Following \citet{Turner+2015} for the non-relativistic solution, and our method for the relativistic jet, we adopt a numerical scheme using a fourth-order Runge-Kutta method in terms of a system of three first order ODEs. 

The following system of equations must be solved for each small angular element $[\theta - \delta\theta/2, \theta + \delta\theta/2)$ of the lobe and shocked shell:
\begin{equation}
\begin{split}
\dot{R}_s(\theta) &= v_s(\theta) \\
\dot{v}_s(\theta) &= \frac{3 (\Gamma_{\rm c} - 1) Q R_s^{\beta - 3}(\theta) \delta\lambda(\theta)}{4 \pi v_s(\theta)\!\; [1 + (\gamma_s v_s/c)^2](\theta)\!\; [\zeta_{\rm s} \gamma_s/\eta_{\rm s}]^2(\theta)\, k \sin\theta \delta\theta} \\
&\quad\quad + \frac{(\beta - 3\Gamma_{\rm c}) v_s^2(\theta)}{2 R_s(\theta)\!\; [1 + (\gamma_s v_s/c)^2](\theta)} \\
\dot{\gamma}_s(\theta) &= \frac{\gamma_s^3(\theta) v_s(\theta) \dot{v}_s(\theta)}{c^2} ,
\end{split}
\label{supersonic system}
\end{equation}
where $k$ and $\beta$ are weakly dependent on $\theta$ as the bow shock at each angle $\theta$ passes over the boundaries between the power-laws approximating the density and temperature profiles at different times.
The differential equations describing the evolution of the shocked shell/lobe in the subsonic and remnant phases are unchanged from those given in \citet{Turner+2015} and \citet{Turner+2018}, with only subtle changes for the weakly supersonic phase due to the omission of the adiabatic index terms compared to the non-relativistic jump conditions. The differential equations for these phases are implemented in {our} {RAiSE} code.

The fraction of the input energy supplied to each small volume element is set to inflate the ellipsoidal shocked shell with the desired axis ratio $A_s$.
Following Equation 22 of \citet{Turner+2015}, the fraction of the energy supplied by the jet kinetic energy to the volume element $\delta V(\theta)$ is given by,
\begin{equation}
\delta\lambda(\theta) = \frac{{\eta_{\rm s}}^{3 - \beta}(\theta)\!\; {\zeta_{\rm s}}^2(\theta) \sin\theta \delta\theta}{\int_0^{\tfrac{\pi}{2}} {\eta_{\rm s}}^{3 - \beta}(\theta')\!\; {\zeta_{\rm s}}^2(\theta') \sin\theta' \delta\theta'} .
\end{equation}
The fraction of energy input to each volume element is held constant upon the commencement of lobe formation (see Section \ref{sec:Two phase fluid}); this permits the shocked shell to elongate as it expands into a steepening ambient gas density profile, rather than being locked into a constant axis ratio.

\subsubsection{Equipartition magnetic field}

The {synchrotron emissivity} from electron populations in the lobe plasma is set by the local magnetic field strength, which is related by a constant of proportionality (equipartition factor $q$; see Section \ref{sec:Particle emissivity model}) to the thermal pressure of the lobe. We therefore need to disentangle the relative contributions of the ram pressure, $p_{r\!\;\!a\!\;\!m}(\theta)$, and thermal pressure, $p(\theta)$, to the momentum flux in Equation \ref{plobe}. This is difficult to achieve directly using the conservation equations, however, the thermal pressure dominates the lobe expansion at very large times in our model. As such, we know the thermal pressure of the lobe at (e.g.) $t=13.8$\,Gyr, as assumed in our code. In practice, the velocity and pressure at this very large time must be calculated assuming expansion in the strong-shock limit, excluding any remnant phase. The thermal component of the pressure at earlier times can then be found by solving, in the reverse time-direction, the differential equation for a lobe with no jet-dominated expansion phase, and again expanding in the strong-shock limit. 

We describe the expansion of this hypothetical lobe, dominated by thermal pressure throughout its evolution, using a recursive differential equation for the four-velocity \citep[cf.][]{KA+1997}; we refer to this as the \emph{thermal} four-velocity. 
The acceleration of the associated shocked shell is written in terms of the thermal four-velocity $w_s$ as:
\begin{equation}
\dot{w}_s(\theta) = \frac{(\beta - 2) w_s(\theta)}{(5 - \beta) t} ,
\label{extra system}
\end{equation}
where we solve for the thermal four-velocity at the first numerical time step $t_0$, $w_s(\theta,t=t_0)$, using the known four-velocity at $t=13.8$\,Gyr. 

The thermal component of the pressure in the relativistic shocked shell cannot exceed that of this hypothetical lobe, however, it may be lower at later times if, for example, the jet switches off. That is, modifying Equation \ref{plobe} for the momentum flux of the shocked shell, we define the thermal component of the pressure as:
\begin{equation}
p(\theta) = k R_s^{-\beta}(\theta) [\min\{ u_s, w_s \} \zeta_s/ \eta_s]^2(\theta) + (k l) R_s^{-\beta}(\theta) ,
\label{thermal}
\end{equation}
where the minimum of the full-model (as derived in Section \ref{sec:Relativistic fluid dynamics}) and thermal velocities is taken to ensure the pressure in remnant lobes (not described by the recursive relationship) is correctly captured. We further assume this pressure represents the (relatively small) thermal component of the momentum flux during the jet expansion phase to provide a smooth transition in emissivity during lobe formation; this agrees well with the hydrodynamic simulation outputs discussed in Section \ref{sec:ANALYSIS OF DYNAMICAL MODEL PREDICTIONS}, though we note that the jet emissivity is not a focus of this work.

\subsection{Dynamics of lobe formation}
\label{sec:Two phase fluid}

The energy supplied by the AGN jet is initially focussed over a small range of angles, $0 \leqslant \theta \leqslant \theta_j$, within the apparent half-opening angle of the jet. Beyond some lobe formation length-scale the energy must distribute across the $2\pi$ steradians of the shocked shell. The source expansion in these two phases is described by the differential equations for the relativistic jet and lobe derived in the previous two sections. In this section we describe the transition between the jet- and lobe-dominated phases based on the relative densities of the jet and ambient medium.

\subsubsection{Two-phase fluid}

We model the radio source expansion as a two-phase fluid; i.e. each angular volume element is assumed to comprise a fraction $\Lambda(t)$ of lobe plasma at any given time $t$. More precisely, the acceleration of the shock front is related to the acceleration of the jet-head, $\dot{v}_{s,j\!\;\!e\!\;\!t}$ (Equation \ref{jet system}), and the acceleration of the shocked shell in the lobe-dominated expansion phase, $\dot{v}_{s,l\!\;\!o\!\;\!b\!\;\!e}$ (Equation \ref{supersonic system}). That is,
\begin{equation}
\dot{v}_s(\theta) = [1 - \Lambda] \dot{v}_{s,j\!\;\!e\!\;\!t} \eta(\theta) + \Lambda \dot{v}_{s,l\!\;\!o\!\;\!b\!\;\!e} (\theta) ,
\label{twophase}
\end{equation}
where $\Lambda$ is the fractional contribution of the lobe plasma to the acceleration of the shocked shell at a given time. The other two coupled ODEs are identical for both fluids and thus do not require any modification.

The time at which lobe formation \emph{commences} provides a potentially useful metric to define the transition from a jet-dominated to a lobe-dominated flow. Such a metric is derived by equating the densities of the jet plasma and ambient medium \citep[e.g.][]{Alexander+2006, Krause+2012}. That is,
\begin{equation}
\mathcal{L}(t) = \frac{\bar{\rho}_j}{\rho_x} = \ \ \frac{\eta_R(R_s(\theta=0, t))}{\bar{\gamma}_j^2}
\label{bigL}
\end{equation}
where $\eta_R(r)$ is defined in Equation \ref{L} and is evaluated for the length of the jet at time $t$. Note that $\mathcal{L} \rightarrow 0$ in the lobe-dominated expansion phase and $\mathcal{L} \rightarrow \infty$ in the jet-dominated phase.

We transform this lobe formation metric to a bounded, monotonically increasing variable, as required in Equation \ref{twophase} (i.e. $\Lambda(t) \in [0,1]$), using a Gaussian distribution as follows:
\begin{equation}
\Lambda(t) = e^{-\mathcal{L}^2(t)/(2 \log 2)} ,
\label{lambda}
\end{equation}
where $\log$ is the natural logarithm.

\subsubsection{Lobe growth and evolution}
\label{sec:Lobe axis ratio evolution}

The two-phase fluid model detailed above considers the evolution of the shocked shell across the transition from a jet-dominated to lobe-dominated flow. In practice, the lobe will initially be confined to a narrow region around the jet before ultimately inflating to its full volume; i.e. an ellipsoid with axis ratio $A(t \rightarrow \infty) \equiv A = A_s^{1/\iota}$, where $A_s$ is the slowly time-varying axis ratio of the shocked shell and $\iota$ is a scaling factor informed by hydrodynamic simulations (see Section \ref{sec:Shock-accelerated plasma lobe}). 

We begin by deriving the volume of the lobe at times shortly after the commencement of lobe formation; we assume the newly forming lobe, as for the jet and jet-head region, is comprised entirely of highly-relativistic plasma. The volume is derived from the total rest-mass of particles injected from the jet, and the volume density of particles in the relativistic plasma, $\rho$. The density is found from the thermal pressure (Equation \ref{thermal}) as follows:
\begin{equation}
\rho_{r\!\;\!e\!\;\!l}(t) = \frac{\Gamma_j p(t)}{h_j c_j^2} \ \ = \frac{4p(t)}{c^2}
\label{cj}
\end{equation}
where the sound speed and adiabatic index are $c_j = c/\sqrt{3}$ and $\Gamma_j = \tfrac{4}{3}$ respectively for a highly-relativistic fluid. The rest-mass of all particles produced by the central black hole that reach the jet-head up to some time $t$ is given by,
\begin{equation}
M(t) = \frac{Q \min\{ t, t_{o\!\;\!n} \}}{(h_j \bar{\gamma}_{j} - 1) c^2} ,
\label{massinj}
\end{equation}
where $t_{o\!\;\!n}$ is the active age of the source, and the Lorentz factor of the bulk flow is specified.
The lobe volume can therefore be written as:
\begin{equation}
V_{r\!\;\!e\!\;\!l}(t) = \frac{M(t)}{\rho_{r\!\;\!e\!\;\!l}(t)} = \frac{Q \min\{ t, t_{o\!\;\!n} \}}{4 p(t) [h_j \bar{\gamma}_{j} - 1]} .
\label{vrel}
\end{equation}
However, this equation only holds shortly after the commencement of lobe formation. At later times, the sound speed (and thus volume) will be greatly reduced as much of the lobe plasma is non-relativistic.

The lobe volume at later times is modelled with a smoothing function based on two upper bounds on the lobe volume: (1) the volume occupied by the injected plasma assuming it is a highly-relativistic fluid (as above); and (2) the maximum volume the lobe can occupy within the shocked gas shell (i.e. for a lobe axis ratio $A(t) \geqslant A_s^{1/\iota}$). That is,
\begin{equation}
V(t) = 1 \bigg/ \left[ \left(\frac{\psi}{V_{r\!\;\!e\!\;\!l}(t) }\right)^\varsigma + \left(\frac{1}{A_s^{-2/\iota} R_s^3(\theta=0, t)}\right)^\varsigma\, \right]^{1/\varsigma} ,
\label{vol}
\end{equation}
where $\psi \lesssim 1$ is the filling factor of the lobe plasma and $\varsigma > 0$, the exponent of this `quadrature' sum, modifies the rate at which the sound speed reduces from its highly-relativistic value; these two parameters are constrained in this work using hydrodynamic simulations. This expression for the lobe volume converges to $V_{r\!\;\!e\!\;\!l}(t)/\psi$ close to the commencement of lobe formation and to the maximum volume the lobe can occupy at late times. 
The time-varying axis ratio of the lobe, $A(t)$, is derived from this volume assuming an ellipsoidal lobe with major axis length $R(\theta = 0,t) = R_s(\theta = 0,t)/b$. That is,
\begin{equation}
A(t) = \left[ \frac{2 \pi R_s^3(\theta=0,t)}{3 b^3 V(t)} \right]^{1/2} .
\label{axisratio}
\end{equation}

The density, and thus sound speed and temperature, of the lobe can be evaluated using the above expression for the lobe volume (Equation \ref{vol}) and the total rest-mass of injected particles in Equation \ref{massinj} to provide a complete description of the dynamics.

\section{PARTICLE EMISSIVITY MODEL}
\label{sec:Particle emissivity model}

The synchrotron emission model in this work is built upon the spatial distribution of Lagrangian tracer particles advected with the fluid flow in relativistic hydrodynamic simulations. 
The particle pressure, density and shock-acceleration histories are used in post-processing \citep[following][]{Yates+2022} to derive the synchrotron emission including adiabatic and radiative losses. 
\citet{Turner+2018a} found that the behaviour of the plasma backflow in the lobes of powerful FR-IIs is highly comparable for vastly different jet powers, axis ratios and ambient gas densities; i.e. the flow is scale invariant. In principle, it should therefore be possible to model the emissivity for radio sources across this broad parameter space by modifying the dynamics of a single set of particles. We present a technique to adapt the dynamics of these Lagrangian particles to match the analytical model rather than that of the hydrodynamic simulation (Section \ref{sec:Lagrangian particle dynamics}). We then summarise the \citet{Turner+2018a} resolved emissivity model including modifications to (e.g.) consider relativistic beaming (Section \ref{sec:Lagrangian particle emissivity}) in order to produce surface brightness images at {radio frequencies} (Section \ref{sec:Surface brightness}).

\subsection{Lagrangian particle dynamics}
\label{sec:Lagrangian particle dynamics}

Lagrangian tracer particles, and parameterisations of their evolutionary histories, are taken from hydrodynamic simulations run by \citet{Yates+2022}. In their PRAiSE model, Lagrangian particles are injected into the lobe plasma approximately every 0.5\,kyr and advected on the Eulerian simulation grid according to the local fluid velocity. The local (to the particle) properties of the fluid are sampled every 10\,kyr.
Regions of shock-acceleration are flagged on the simulation grid using the flagging scheme described by \citet{Mignone+2012} for several different shock thresholds; we assume a particle shock threshold of 5 following \citet{Yates+2022}. The result for a typical simulated source age of 35\,Myr is approximately 50\,000 particles with pressure, density and shock-acceleration histories. 
In this work, we only consider temporal snapshots in which the lobe and shocked shell have stabilised into their late-time self-similar expansion phase; however, we approximate the emissivity in the jet-dominated and remnant phases by selectively darkening jet and lobe particles (see Section \ref{sec:Lagrangian particle emissivity}).

\subsubsection{Shock-acceleration time}

The time the Lagrangian particles are last shock-accelerated needs to be updated to consider both: (1) the difference in source age between the analytical model, $t$, and hydrodynamic simulation, $\hat{t}$; and (2) the active age of the source, $t_{on}$, in the case of remnants. 

For the $i$-th Lagrangian particle, the dynamical model time $t_{a\!\;\!c\!\;\!c, i}$ at which the particle is taken to have been most recently accelerated is related to that of the hydrodynamic simulation, $\hat{t}_{a\!\;\!c\!\;\!c, i}$, as follows:
\begin{equation}
t_{a\!\;\!c\!\;\!c, i} = \hat{t}_{a\!\;\!c\!\;\!c, i} \left( \frac{t}{\hat{t}} \right) \min\{1, \frac{t_{o\!\;\!n}}{t} \} ,
\end{equation}
where the final shock-acceleration time in the hydrodynamic simulation occurs at the simulated source age (i.e. $\max\{..., \hat{t}_{a\!\;\!c\!\;\!c, j}, ... \} = \hat{t}$).

In the case of remnants with recently switched off jets, we make use of the same Lagrangian particles under the assumption that their locations remain approximately constant for a short time after the cessation of jet activity; this assumption is made to ensure that previously shock-accelerated particles are present throughout the lobe, including in the jet-head region.

\subsubsection{Particle locations}
\label{sec:Particle locations}

The location of the Lagrangian particles is required to produce surface brightness maps and modify the emissivity during lobe formation. The location of the $i$-th particle along the jet axis is given by,
\begin{equation}
z_i = \hat{z}_i \frac{R_s(\theta = 0)}{b \max\{...,\hat{z}_j,... \}} ,
\end{equation}
where $\hat{z}_i$ is the location of the $i$-th particle in the hydrodynamic simulation, and $\max\{...,\hat{z}_j,... \}$ is the length of the jet axis (i.e. maximum distance reached by a lobe particle of any index $j$). The lobe length from the dynamical model is $R(\theta = 0) = R_s(\theta = 0)/b$ as described in Section \ref{sec:Shock-accelerated plasma lobe}. The location of the Lagrangian particles along the $x$- and $y$-axes are similarly updated to match the analytical dynamical model (replacing all $z$ terms in the above equation by $x$ and $y$ respectively) but scaled to the width of the late-time lobe instead of its length; i.e. $A_s^{1/\iota} R_s(\theta = 0)/b$

\subsubsection{Electron packet volume}

We derive the volume associated with each of the Lagrangian particles, $\delta\hat{V}_i$, by calculating the Voronoi tessellation of all particles using their Cartesian coordinates, $(\hat{x}_i, \hat{y}_i, \hat{z}_i)$. The volume of each Voronoi region is found by calculating the corresponding convex hull \citep[cf.][]{Yates+2022}. Particles along the edge of the lobe will have partially (or completely) unconstrained volumes that tend towards infinity. As a result, we restrict the maximum volume of any particle to the $n$-th percentile volume such that the total volume of \emph{lobe} particles is equal to that of the lobe cavity for the hydrodynamic simulations used in this work. 

Particles that are entrained by the shocked gas shell or ambient medium must be flagged as these will rapidly lose their energy through thermal interactions rather than synchrotron emission. The hydrodynamic simulations run by \citet{Yates+2022} include a passive tracer field which measures the fraction of the fluid in each grid cell that originates from the jet injection cells, $\hat{\psi}$ (i.e. the filling factor); only particles with $\hat{\psi}_i > 10^{-4}$ are included in the lobe. The total volume occupied by lobe particles is thus given by,
\begin{equation}
\hat{V} = \sum_i \begin{cases} \min \{ \delta \hat{V}_i, \text{\%tile}(\{..., \delta \hat{V}_j, ...\}, n) \} ,& \hat{\psi}_i > 10^{-4} \\ 0, & \text{otherwise} \end{cases} ,
\end{equation}
where $\delta \hat{V}_i$ is the volume associated with the $i$-th Lagrangian particle injected into the lobe (i.e. with $t_{a\!\;\!c\!\;\!c, i} \leqslant t$), and $\hat{\psi}_i$ is the corresponding lobe tracer field. Here, we define $\text{\%tile}(\{ ...\}, n)$ as the $n$-th percentile of some scalar quantity, in this case, the particle volume. For the hydrodynamic simulations used in this work, we find the 95th percentile yields a total particle volume equal to that of the lobe cavity.

The volume occupied by each Lagrangian particle includes contributions from the synchrotron-emitting electron population injected by the jet in addition to the ambient particles (as described by the filling factor). The volume associated with each Lagrangian particle, that contributes to synchrotron emission, is updated for the dynamics of the analytical model using the following equation:
\begin{equation}
\delta V_i = \hat{\psi}_i \delta \hat{V}_i  \left(\frac{2V}{\hat{V}} \right) ,
\label{corrvol}
\end{equation}
where $V$ is the volume of a single lobe in the analytical model (Equation \ref{vol}); the factor of two is present because the hydrodynamic simulations of \citet{Yates+2022} models both lobes.

\subsubsection{Pressure evolutionary histories}
\label{sec:Pressure evolutionary histories}

The evolutionary histories of the pressure associated with the Lagrangian particles are required to derive the synchrotron emissivity including adiabatic and radiative loss mechanisms. The pressure of the particles therefore needs to be considered (at minimum) at both the time of shock acceleration in addition to the source age (when emission occurs). As a result, the simulated pressure evolutionary histories for each particle, $\hat{p}_i$, need to be adapted to consider: (1) the spatial distribution of pressure across the lobe; and (2) the change in this pressure distribution since the shock-acceleration time.

In the analytical model, the lobe pressure (see Equation \ref{thermal}) is expected to be spatially smooth with a gradient from the jet-head (higher pressure\footnote{In remnants the pressure along the major axis may be lower than that along the minor axis.}) to the minor axis of the lobe (lower pressure). The pressure along the minor axis of the lobe provides a robust absolute scaling of the hydrodynamic simulation particle pressure to that of the analytical model; by contrast, the jet-head region is not captured in detail by the analytical model. That is,\footnote{The superscript $(d)$ on the updated pressure of the $i$-th particle, $p_i^{(d)}$, represents the number of corrections so far applied to that Lagrangian particle.}
\begin{equation}
{p}_i^{(1)} = \hat{p}_i \left(\frac{p(\theta = \tfrac{\pi}{2})}{\hat{p}_{m\!\;\!i\!\;\!n\!\;\!o\!\;\!r}} \right) ,
\end{equation}
where $p(\theta = \tfrac{\pi}{2})$ is the thermal pressure of the lobe along the minor axis from the analytical model, and $\hat{p}_{m\!\;\!i\!\;\!n\!\;\!o\!\;\!r}$ is the median value for hydrodynamic simulation particles along that axis with a lobe tracer field value $\hat{\psi} > 10^{-4}$.

Meanwhile, the pressure gradient along the jet axis is modelled through a monotonic function $f(z)$, based on the ratio of the jet-head and minor axis pressures in the analytical model and hydrodynamic simulation. However, the pressure in the jet-head region of the hydrodynamic simulations is increased by local-scale physics not included in the analytical model; to enable a direct comparison, this pressure ratio (jet-head/minor axis) is instead estimated from the bow shock velocities for the strong-shock limit expansion of the simulated sources, at least when taking particles from simulations of high-powered sources as in this work. That is, the pressure ratio is approximately related to the axis ratio of the shocked shell as $\hat{p}_{m\!\;\!a\!\;\!j\!\;\!o\!\;\!r}/\hat{p}_{m\!\;\!i\!\;\!n\!\;\!o\!\;\!r} \approx \hat{A}_s^2$ \citep[cf.][]{KA+1997}.
The pressure of a Lagrangian particle at some location ${z}_i$ along the jet axis is updated for the changed pressure gradient as follows:
\begin{equation}
\begin{split}
p_{i}^{(2)} &= {p}_i^{(1)} \left[\frac{p(\theta=0)/p(\theta=\tfrac{\pi}{2})}{\hat{A}_s^2} - 1 \right] f\!\left(\frac{|z_i|}{R(\theta=0)}\right) + {p}_i^{(1)} ,
\end{split}
\end{equation}
where $|z_i|/R(\theta=0)$ is the fractional distance of the $i$-th particle along the jet axis. The function $f(z)$ describes the relationship between the location of the particle and the expected fractional change in pressure; we assume a linear function which should be sufficient for moderate changes in the pressure ratio.

Changes in the evolutionary history of the pressure between the analytical model and hydrodynamic simulations is largely driven by the shape of the host gas density profile; for example, in the strong-shock limit the lobe pressure is $p \propto t^{(-4 - \beta)/(5 - \beta)}$ \citep[e.g. Equation 20 of][]{KA+1997}. The evolutionary history of the pressure associated with each particle is modelled as a power-law of the form:
\begin{equation}
\hat{p}_{a\!\;\!c\!\;\!c, i} = \hat{p}_i \left(\frac{\hat{t}_{a\!\;\!c\!\;\!c, i}}{\hat{t}} \right)^{\hat{a}_{i}} ,
\end{equation}
where $\hat{p}_{a\!\;\!c\!\;\!c, i}$ is the pressure of the $i$-th particle at its shock-acceleration time $\hat{t}_{a\!\;\!c\!\;\!c, i}$, and $\hat{a}_{i}$ is an exponent representing the pressure evolution of that particle. This evolutionary history is updated for the environment of the analytical model by considering the ambient gas density encountered by the lobe as a function of time, both for the hydrodynamic simulation and for the analytical model. That is,
\begin{equation}
\begin{split}
&\hat{\rho}_{a\!\;\!c\!\;\!c, i} = \hat{\rho}_i \left(\frac{\hat{t}_{a\!\;\!c\!\;\!c, i}}{\hat{t}} \right)^{\hat{\varepsilon}_{i}} \ \ \ \text{and} \\
&\rho_{a\!\;\!c\!\;\!c, i} = \rho_i \left(\frac{t_{a\!\;\!c\!\;\!c, i}}{t} \right)^{\varepsilon_{i}} ,
\end{split}
\end{equation}
where $\hat{\varepsilon}_{i}$ and ${\varepsilon}_{i}$ are the exponents of the gas density--time profiles for the hydrodynamic simulation and analytical model respectively. The exponents for the hydrodynamic simulation density profile are pre-calculated and stored as a property of the Lagrangian particles. 
The pressure associated with the $i$-th Lagrangian particle, at the time of shock-acceleration, therefore has its pressure updated for the dynamics of the analytical model as:
\begin{equation}
{p}_{a\!\;\!c\!\;\!c, i}^{(3)} = {p}_i^{(2)} \left(\frac{t_{a\!\;\!c\!\;\!c, i}}{t} \right)^{\hat{a}_{i} + \varepsilon_i - \hat{\varepsilon}_{i}} .
\end{equation}

The corrected thermal pressure at the source age and shock-acceleration time are hereafter written as $p_{i}\equiv p_{i}^{(2)}$ and ${p}_{a\!\;\!c\!\;\!c, i} \equiv {p}_{a\!\;\!c\!\;\!c, i}^{(3)}$ respectively.

\subsection{Lagrangian particle emissivity}
\label{sec:Lagrangian particle emissivity}

The {synchrotron emissivity model used in this work follows the method presented by \citet{Turner+2018a}. The radiative losses predicted by this model} are excluded from the energy budget considered in our dynamical model; i.e. we implicitly assume that the radiative losses are sufficiently small to not appreciably affect the source dynamics. \citet{Hardcastle+2018} shows this is a reasonable assumption for most sources, though the inverse-Compton emission can become a significant factor in the oldest sources at high-redshift.

\subsubsection{Synchrotron emissivity}
\label{sec:Synchrotron emissivity}

The luminosity at rest-frequency $\nu$ due to synchrotron-emission from electrons in the volume element $\delta V_i$ associated with the $i$-th Lagrangian particle (Equation \ref{corrvol}) is given by Equation 9 of \citet{Turner+2018a}:
\begin{equation}
\begin{split}
\delta L_{s\!\;\!y\!\;\!n, i}&(\nu, s, z) = K(s) \nu^{(1 - s)/2} \frac{q^{(s + 1)/4}}{(q + 1)^{(s + 5)/4}} \delta V_i \\
&\!\!\!\!\times p_{i}^{(s + 5)/4}  \left[\frac{p_{a\!\;\!c\!\;\!c, i}}{p_{i}} \right]^{1 - 4/(3\Gamma_{\rm c})} \left[\frac{\gamma_{a\!\;\!c\!\;\!c, i}(\nu, z)}{\gamma_{i}(\nu)} \right]^{2 - s}  ,
\end{split}
\end{equation}
where $K(s)$ is the source specific constant defined in Equation 5 of \citet{Turner+2018a}, and $q \equiv u_B / u_e$ is the ratio of the energy density in the magnetic field to that in the synchrotron-emitting particles \citep[equipartition factor;][]{Turner+2018b}. The thermal pressure of the $i$-th Lagrangian particle updated for the dynamics of the analytical model at the source age $t$ is defined as $p_{i}$, and the pressure of that particle at the time of shock-acceleration is defined as $p_{a\!\;\!c\!\;\!c, i}$ (see Section \ref{sec:Pressure evolutionary histories}).

The Lorentz factor of the $i$-th Lagrangian particle at the source age, when the synchrotron radiation is emitted, is given by,
\begin{equation}
\gamma_i(\nu) = \left[ \frac{2\pi m_e \nu}{3e} \sqrt{\frac{(\Gamma_c - 1)(q + 1)}{2 \mu_0 p_i q}}\, \right]^{1/2} ,
\end{equation}
where $e$ and $m_e$ are the charge and mass of the electron respectively, and $\mu_0$ is the vacuum permeability. The Lorentz factor at the shock-acceleration time is found following the method of \citet{KDA+1997} assuming a single power-law for the evolutionary history of the pressure associated with each particle. That is,
\begin{equation}
\gamma_{a\!\;\!c\!\;\!c, i}(\nu, z) = \frac{\gamma_i(\nu) t_{a\!\;\!c\!\;\!c, i}^{a_{i}/(3\Gamma_c)}}{t^{a_{i}/(3\Gamma_c)} - a_2(t, t_{a\!\;\!c\!\;\!c, i}) \gamma_i(\nu)} ,
\label{gacc}
\end{equation}
where the constant $a_2(t, t_{a\!\;\!c\!\;\!c, i})$ is given by,
\begin{equation}
\begin{split}
a_2(t, &t_{a\!\;\!c\!\;\!c, i}) = \frac{4 \sigma_T}{3 m_e c} \left[ \frac{p_{a\!\;\!c\!\;\!c, i} q t_{a\!\;\!c\!\;\!c, i}^{-a_{i}}}{a_3 (\Gamma_c - 1)(q + 1)}(t^{a_3} - t_{a\!\;\!c\!\;\!c, i}^{a_3}) \right. \\
&+ \left. \frac{u_{c0} (1 + z)^4}{a_4} (t^{a_4} - t_{a\!\;\!c\!\;\!c, i}^{a_4}) \right],
\end{split}
\label{a2}
\end{equation}
where $\sigma_T$ is the Thompson electron scattering cross-section, $c$ is the speed of light, and $u_{c0} = 4.00\times 10^{-14}\rm\, J\, m^{-3}$ is the energy density of the cosmic microwave background radiation at the present epoch \citep[e.g.][]{KDA+1997}. The exponents $a_3$ and $a_4$ are defined in terms of the evolutionary history of the pressure as $a_3 = 1 + a_{i}(1 + 1/3 \Gamma_c)$ and $a_4 = 1 + a_{i}/3 \Gamma_c$.

\subsubsection{Doppler de-boosted jet}

Lagrangian particles in the jet have their synchrotron emission relativistically beamed along the jet axis due to their highly-relativistic velocities ($|v| \gg 0.1c$). 
The brightness of these particles in the observer frame is related to their rest-frame emissivity (as calculated in {Section \ref{sec:Synchrotron emissivity}) }as follows:
\begin{equation}
\delta L_{i,o}^{(1)} (\nu_o) = \delta L_{i} (\nu) \left[ \frac{1}{\gamma_i(1 - \boldsymbol{v}_i \cdot \boldsymbol{\hat{n}}/c) } \right]^{(s + 3)/2} ,
\end{equation}
where $\boldsymbol{v}_i$ is the velocity vector of the $i$-th particle, $\gamma_i$ is the corresponding Lorentz factor, and $\boldsymbol{\hat{n}}$ is the observing normal (i.e. $\boldsymbol{v}_i \cdot \boldsymbol{\hat{n}} = 0$ for jet particles with a radio galaxy in the plane of the sky). Here, the rest-frame frequency, $\nu$, is related to the observer-frame frequency, $\nu_o$, as $\nu = (1 + z)\nu_o$.

The emissivity of the jet particles must further be modified to consider a remnant phase after the cessation of jet activity; the jet particles need to be distinguished from the particles comprising the lobe to exclude their contribution to the radio source emissivity. The emissivity of the Lagrangian particles is parameterised as:
\begin{equation}
\delta L_{i,o}^{(2)}(\nu_o) =\delta L_{i,o}^{(1)}(\nu_o) \begin{cases}
1, & t \leqslant t_{o\!\;\!n} \\
1 - \min \{|\boldsymbol{v}_i \cdot \boldsymbol{\hat{z}}|/c_j, 1\} , & t > t_{o\!\;\!n}
\end{cases}
\label{remjet}
\end{equation}
where $\boldsymbol{\hat{z}}$ is a unit vector along the jet axis and $c_j$ is the sound speed of a highly-relativistic fluid; i.e. the argument $\min \{|\boldsymbol{v}_i \cdot \boldsymbol{\hat{z}}|/c_j, 1\}$ is arbitrarily defined to give a value of unity for particles with a velocity component along the jet axis $|v_z| \geqslant c/\sqrt{3}$. This function tapers off linearly to zero for particles with lower velocities. As a result, the emissivity of the highly-relativistic jet particles is set to zero in the remnant phase, whilst the brightness of the low-velocity lobe particles is unaffected.

\subsubsection{Lobe formation emissivity}

The {synchrotron emissivity calculated from the Lagrangian particles (taken at late-time temporal snapshots; Section \ref{sec:Lagrangian particle dynamics}) needs} to consider the extent reached by shock-accelerated plasma during lobe formation. In Section \ref{sec:Particle locations}, we scaled the locations of these particles based on the dimensions of the shocked shell in the analytical model\footnote{We choose not scale the particle locations directly to the dimensions of the lobe as it has a strongly time-varying axis ratio; with this assumption, the shape of the jet and jet-head region would be highly variable.}. However, the axis ratio of the lobe is much greater during the lobe formation phase than at late-times; i.e. $A(t) \gg A(t\rightarrow \infty) = A_s^{1/\iota}$. Scaled hydrodynamic simulation particles (from the late-time snapshots) located outside the lobe of the analytical model are flagged and have their emissivity set to zero. For the $i$-th particle at radius $r_i = \sqrt{x_i^2 + y_i^2 + z_i^2}$ and angle $\theta_i = \arctan(\sqrt{x_i^2 + y_i^2}/z_i)$ the emissivity is as follows:
\begin{equation}
\delta L_{i,o}^{(3)}(\nu_o) = \delta L_{i,o}^{(2)}(\nu_o) \begin{cases}
1,  & r_i < r_{i,c\!\;\!r\!\;\!i\!\;\!t}\\
\min \{|\boldsymbol{v}_i \cdot \boldsymbol{\hat{z}}|/c_j, 1\} , & \text{otherwise}
\end{cases},
\end{equation}
where $r_{i,c\!\;\!r\!\;\!i\!\;\!t} = \eta(\theta_i) R_s(\theta_i)/[b \eta_s(\theta_i)]$; here, $\eta(\theta_i)$ is evaluated using the present-time lobe axis ratio, $A(t)$, derived in Equation \ref{axisratio}, and $\eta_s(\theta_i)$ is in terms of the axis ratio of the shocked shell, $A_s$. The emissivity of particles outside the lobe are set close to zero except for those with a highly-relativistic velocity component along the jet axis, using the argument discussed in the previous section; this condition is required to ensure the jet particles are visible prior to the commencement of lobe formation.

\subsection{Surface brightness images}
\label{sec:Surface brightness}

High-resolution surface brightness maps require more than a single temporal snapshot of particles for the hydrodynamic simulation to produce detailed images. This is easily achieved as our framework in Section \ref{sec:Lagrangian particle dynamics} scales any of the late-time snapshots of Lagrangian particles to the source dynamics at the present time. We therefore combine the particle locations, $(x_i, y_i, z_i)$, and observer-frame emissivities, $\delta L_{i,o}$, of $m$ temporal snapshots to increase the effective number of particles contributing to a given surface brightness image. The properties of these particles are stored in a compressed file format, \texttt{hdf5}, that is sufficiently small to be included in online software repositories; however, this restricts the number of unique snapshots to the order of a few hundred. We rotate the locations of the particles in these snapshots along the jet axis, and mirror the particles between the two simulated lobes, to obtain a greater number snapshots whilst smoothing over any asymmetries in the simulations (which are not captured in the analytical model). The highest resolution images in this work are therefore able to combine 2048 snapshots, and include 115 million particles.

The lobes of each mock radio galaxy are divided into an $n \times n \times n$ grid of cubic pixels; each of the particles is assigned to its spatially coincident pixel based on the adapted $(x_i, y_i, z_i)$ locations. The pixel size is chosen to be sufficiently large such that any line of sight through the source is associated with multiple particles. For the highest resolution images in this work we assume $n = 512$ pixels.
The two-dimensional surface brightness is simply calculated by summing the emissivity from every cell along the depth of the source, assuming the lobe plasma and ambient medium in front of the source is optically thin. The surface brightness images we present in this work assume the radio source is located in the plane of the sky.

\section{ANALYSIS OF MODEL PREDICTIONS}
\label{sec:ANALYSIS OF DYNAMICAL MODEL PREDICTIONS}

In this section, we compare the dynamics of {our new} RAiSE model (described in Section \ref{sec:ANALYTIC DYNAMICAL MODEL}) to the results of hydrodynamic simulations run using the PLUTO code \citep{Mignone+2007}. We describe both the existing simulations run by \citet{Yates+2022} for powerful FR-IIs (Section \ref{sec:Hydrodynamic Simulations}) and a new simulation of a low-powered, lobed FR-I performed for this work (Section \ref{sec:Dynamics of low-powered FR-I}).
We validate, and calibrate, the analytical model by comparing the predicted evolution of the jet-head/lobe length and axis ratio throughout the source lifetime (Sections \ref{Sec:Free parameters in RAiSE} and \ref{sec:Dynamics of low-powered FR-I}), and verify the spatial distribution of relevant dynamical quantities by comparing the modelled surface brightness images (Section \ref{sec:Comparison of brightness images}).

\subsection{Hydrodynamic simulations}
\label{sec:Hydrodynamic Simulations}

The {new} RAiSE emissivity model is based on Lagrangian particles from the hydrodynamic simulations of \citet{Yates+2022}. These simulations are identical to those studied by \citet{Yates+2021} but additionally include particles to incorporate radiative loss mechanisms as presented in \citet{Turner+2018a}. In this paper, we focus exclusively on differences between the dynamics of the models and any resulting differences in spatial brightness distributions; \citet{Yates+2022} provide a verification of the RAiSE integrated luminosity through a comparison with an alternative particle-based brightness calculation.

The two hydrodynamic simulations of \citet{Yates+2022} both consider a high-powered FR-II jet ($Q_{t\!\;\!o\!\;\!t} = 10^{38.8}$\,W, or $Q = 10^{38.5}$\,W) with Lorentz factor $\gamma_j = 5$ and half-opening angle $\theta_{j0} = 10$ degrees (for the inner, un-collimated section of the jet), expanding into a cluster environment at redshift $z = 0.05$. The environment is modelled as a spherically symmetric King profile with core density of $\rho_c = 2.41\times10^{-24}$\,kg\,m$^{-3}$, core radius of $r_c = 144$\,kpc, and slope described by the coefficient $\beta' = 0.38$ \citep[for details, see][]{Yates+2021}. The simulations either have the jets expand outwards from the cluster centre or expanding radially outwards (inwards) for the jet (counterjet) from a cluster radius of $r = r_c$. In this work, we will only consider the cluster-centred jet as the analytic theory underpinning RAiSE, and other analytical models, assumes a spherically symmetric cluster environment. {We investigate source expansion in more representative gas density profiles informed by {X-ray observations of local galaxy clusters} in Section \ref{sec:Jet contribution to evolutionary history} using the then calibrated RAiSE model.}

The emissivity calculations in \citet{Yates+2022} are based on approximately 50\,000 Lagrangian particles viewed at 3526 linearly-spaced temporal snapshots up to an age of 35.1\,Myr. However, early in the source evolutionary history, only a fraction of these particles have been shock-accelerated resulting in poor resolution surface brightness images (see Section \ref{sec:Comparison of brightness images}). The hydrodynamic simulation could of course be rerun with an increased injection rate of Lagrangian particles into the jets to ensure sufficient resolution at earlier source ages.  

The source age, power and bulk velocity of the jet are equivalent parameters between RAiSE and the hydrodynamic simulation, whilst the gas density profile is approximated by a series of 64 power-laws radiating outwards from the active nucleus. However, several of the model parameters in RAiSE are not specified explicitly in the hydrodynamic simulation, and instead must be constrained by calibrating model outputs.

\subsection{Free parameters in RAiSE} 
\label{Sec:Free parameters in RAiSE} 

The {new} RAiSE dynamical model has six model parameters that have no direct counterparts in the PLUTO hydrodynamic simulations; however, these can be readily constrained from the simulation outputs.
These parameters are: the thickness of the shocked gas shell along the jet axis, $b = R_s(\theta = 0)/R(\theta = 0)$; the relative axis ratios of the late-time lobe and shocked shells, $\iota = \log{A_s} / \log{ A}$ (both Section \ref{sec:Shock-accelerated plasma lobe}); the ratio of the radius of jet spine to jet sheath, $a_{*}$ (Section \ref{sec:Jet spine and sheath}); the apparent half-opening angle of the jet, $\theta_j$ (Section \ref{sec:Relativistic hydrodynamic equations}); the lobe filling factor, $\psi$; and the rate at which the sound speed decreases from its highly-relativistic value, $\varsigma$ (both Section \ref{sec:Lobe axis ratio evolution}).

\begin{table}
\begin{center}
\newcolumntype{L}{>{\raggedright\arraybackslash}m{120pt}}
\caption[]{Optimised values of the free parameters in the RAiSE dynamical model constrained using the hydrodynamic simulations of \citet{Yates+2022}.}
\label{tab:calibparams}
\renewcommand{\arraystretch}{1.1}
\setlength{\tabcolsep}{6pt}
\begin{tabular}{cccL}
\hline\hline
Parameter&Value&&Description \\
\hline
$b$&1.07&&Ratio of shocked shell to lobe radius along jet axis\\
$\iota$&0.59&&Axis ratio of shocked shell to lobe in form $A_s = A^\iota$\\
$a_{*}$&0.23&&Ratio of the radius of jet spine to jet sheath\\
$\theta_j$&0.67&&Apparent jet half-opening angle at lobe formation scale ($^\circ$)\\
$\psi$&0.55&&Lobe filling factor\\
$\varsigma$&0.34&&Sound speed rate of decrease from highly-relativistic value\\
\hline
\end{tabular}
\end{center}
\end{table}

\subsubsection{Thickness of shocked gas shell}

The relative shapes of the lobe and shocked shell in the RAiSE dynamical model are described by two variables that can be constrained based on late-time temporal snapshots from the hydrodynamic simulations; i.e. we compare their axis ratios after the lobes are fully formed but at a time that is still well described by the shock-shock supersonic expansion limit (i.e. self-similar expansion). 
The thickness of the shocked gas shell along the jet axis is measured at several temporal snapshots from the hydrodynamic simulation (to mitigate noise due to the non-linear behaviour of the simulated fluids on small-scales) as $b = R_s(\theta = 0)/R(\theta = 0)$. The relative axis ratios of the late-time lobe and shocked shells, $A_s = A^\iota$, is then constrained from the simulations as:
\begin{equation}
\iota = \frac{\log{R_s(\theta = 0)} - \log{R_s(\theta = \tfrac{\pi}{2})}}{\log{R(\theta = 0)} - \log{R(\theta = \tfrac{\pi}{2})}} ,
\end{equation}
where the size of the shocked shell must be measured from the grid-based hydrodynamic simulation outputs, in particular the pressure, as the particles only trace the synchrotron-emitting lobe regions. The chosen values for these two parameters are shown in Table \ref{tab:calibparams}.

\subsubsection{Radius of the jet spine}

The radius of the jet spine is neither an intrinsic parameter of hydrodynamic simulations nor readily extracted from their outputs; however, we can relate this quantity to the easily measurable advance speed of the jet-head during the jet-dominated expansion phase. 
Specifically, examining Equation \ref{vs}, we see that the spatially-averaged bulk velocity of the jet plasma is exactly equal that of the jet-head advance speed early in the jet-dominated expansion phase; i.e. $v_s \rightarrow \bar{v}_j$. We can therefore rearrange Equation \ref{jetvelocity} to obtain an expression for the ratio of the radius of jet spine to sheath in terms of the jet-head advance speed and the Lorentz factor of the jet. That is,
\begin{equation}
a_{*} = \bigg[ \frac{v_s^2}{(\gamma_j^2 - 1)(c^2 - v_s^2)} \bigg]^{1/4} \ \ = 0.23 ,
\end{equation}
where the right-hand equality is obtained by evaluating the expression for the simulated Lorentz factor of $\gamma_j = 5$ and the advance speed measured from the hydrodynamic simulations of \citet{Yates+2022}; i.e. $v_s = 0.252c$. The jet-head advance speed will of course be simulation dependent, however, we assume the internal structure of the jet remains comparable for any `light' relativistic jet. We discuss the case of a `heavy' non-relativistic jet in detail in Section \ref{sec:Dynamics of low-powered FR-I}.

\begin{figure*}
\begin{center}
\includegraphics[width=0.42\textwidth,trim={15 15 40 40},clip]{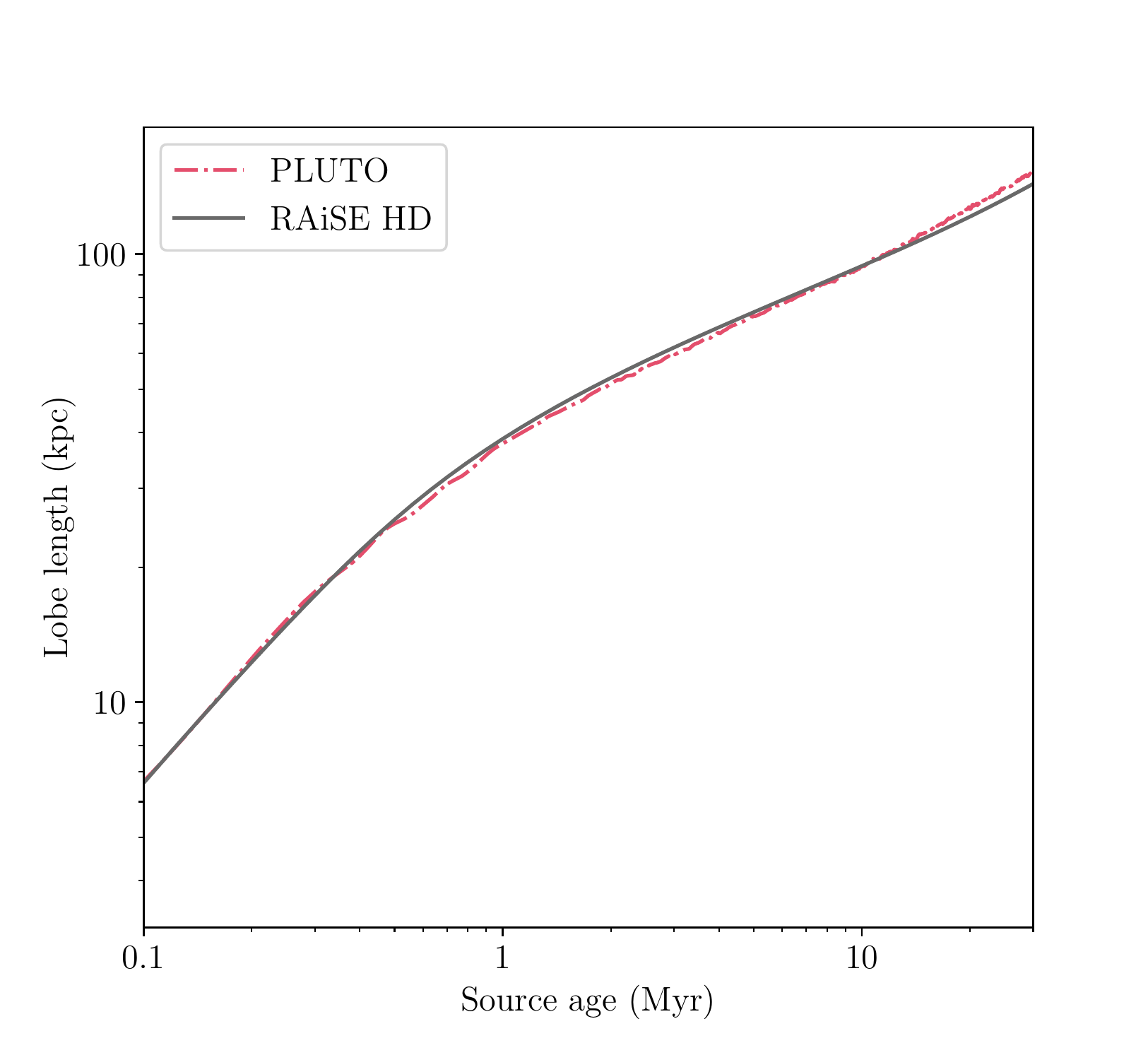} \quad\quad \includegraphics[width=0.42\textwidth,trim={15 15 40 40},clip]{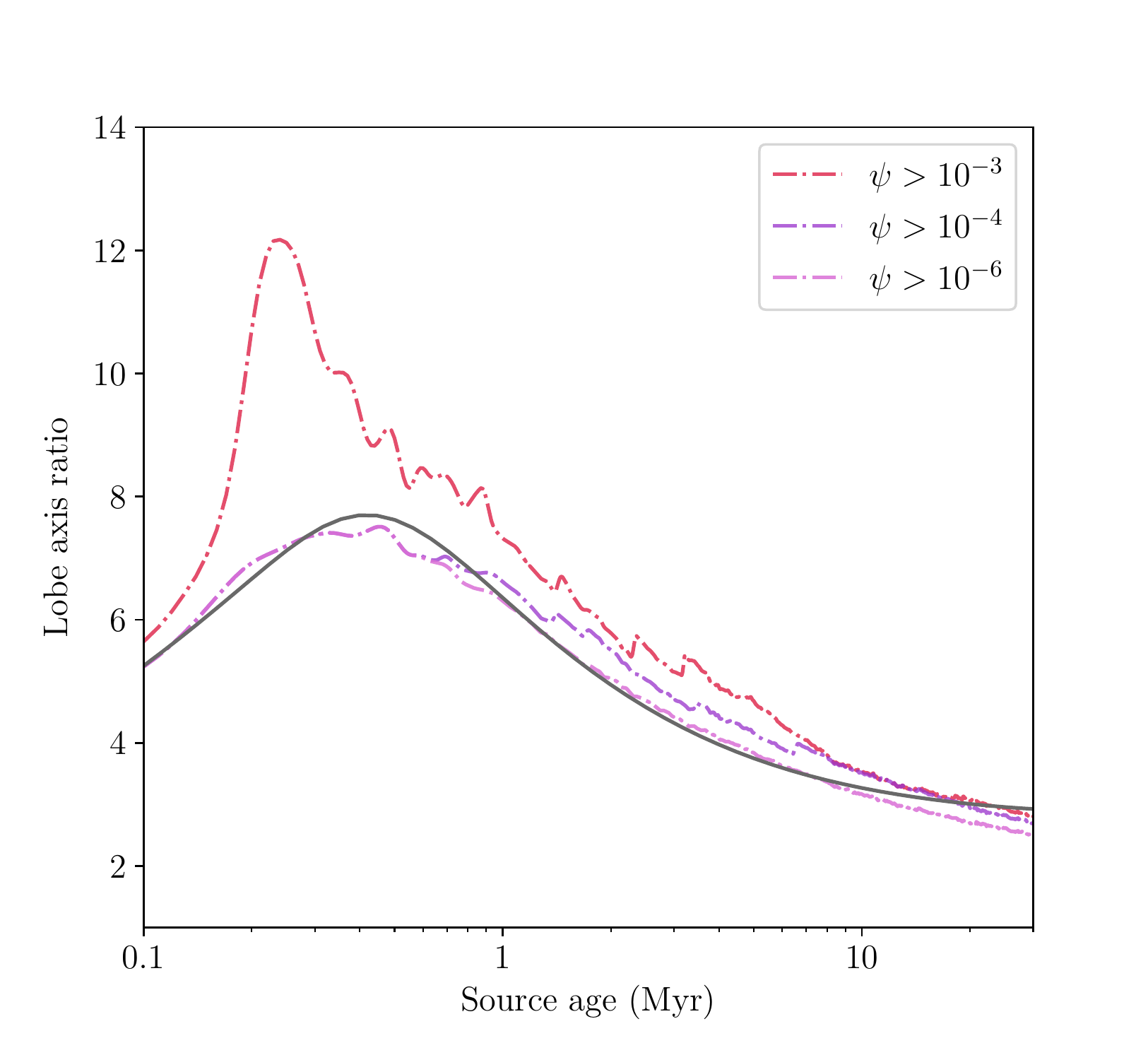}
\end{center}
\caption[]{Calibration of three of the remaining free parameters in the RAiSE dynamical model based on the jet-head/lobe length expansion (left) and axis ratio evolution (right) measured from the (PLUTO) hydrodynamic simulation of a Fanaroff-Riley Type-II radio source by \citet{Yates+2022}. The axis ratio in the hydrodynamic simulation is measured for three different critical jet tracer values (i.e. minimum value for lobe filling factor $\bar{\psi}$).}
\label{fig:calib}
\end{figure*}

\subsubsection{Apparent jet opening angle}

The apparent half-opening angle of the jet is defined at the cessation of the jet-dominated expansion phase; i.e. at the time the governing equations for the acceleration change. Bounds can be placed on this parameter based on the apparent opening angle in the jet-dominated (upper bound) and late-time evolutionary phases (lower bound), however, the `cessation of the jet-dominated expansion phase' cannot be exactly quantified in a hydrodynamic simulation. We instead compare the predicted evolution of the lobe length in RAiSE and the hydrodynamic simulation of \citet{Yates+2022} for a range of apparent half-opening angles; i.e. we choose the apparent half-opening angle that changes the governing different equations for the acceleration at the correct time to match the expected late-time evolution.

The residual sum-of-squares (RSS) between the lobe length of RAiSE and (PLUTO) hydrodynamic simulation for each trial value of the apparent opening angle is as follows:
\begin{equation}
\begin{split}
\text{RSS} &= \frac{1}{N} \sum_{n = 1}^N \big[\log_{10}R(\theta = 0, t_n) - \log_{10}\max\{...,\hat{z}_j(t_n),... \} \big]^2 ,
\end{split}
\label{RRS}
\end{equation}
where $R(\theta = 0, t_n)$ is the RAiSE lobe length (assuming the lobe thickness parameters fitted above) predicted for source age $t_n$, corresponding to the $n$-th (of $N$) temporal snapshot from the hydrodynamic simulation. The lobe length measured from the $n$-th snapshot of hydrodynamic simulation particles is $\max\{...,\hat{z}_j(t_n),... \}$. 
The RAiSE model prediction for the expansion of the jet-head/lobe for the best fit value of the apparent half-opening angle (and the jet and lobe thickness parameters discussed above) is shown in Figure \ref{fig:calib}(left). 

The opening angle of the inner, un-collimated section of the jet, $\theta_{j0}$, is known to scale with the late-time axis ratio of the lobe \citep[e.g.][their Equation 38]{KA+1997}; we assume that the apparent half-opening angle of the jet similarly scales with the axis ratio\footnote{The opening angle of the inner jet is linearly proportional to the cross-sectional radius of the collimated jet, and is therefore linearly proportional to its apparent half-opening angle.}. We must therefore modify the apparent half-opening angle fitted for the jet of \citet{Yates+2022} when considering different late-time lobe axis ratios to their value of $A(t\rightarrow \infty) = 2.83$. That is,
\begin{equation}
\theta_j(A) = \left[\frac{2.83}{A} \right]\!\;\! \theta_{j}(A = 2.83) ,
\label{axisopening}
\end{equation}
where the best fit value for the apparent jet half-opening angle for the hydrodynamic simulation is $\theta_j(A = 2.83) = 0.67$ degrees. 

\subsubsection{Lobe filling factor and sound speed}
\label{sec:Lobe filling factor and sound speed}

The physics of the forming lobe, after the commencement of lobe formation, are described by two free parameters: the (spatially-averaged) lobe filling factor, $\psi$, and the rate at which the sound speed decreases from its highly-relativistic value, $\varsigma$. The filling factor of the lobe plasma is reported for each grid cell (and associated particles) of the \citet{Yates+2022} hydrodynamic simulations; however, deriving an average filling factor across the lobe (as needed to compare with the analytical model) is highly dependent on the critical value of the filling factor used to define the boundary between the lobe and shocked shell. The corresponding evolution in the lobe axis ratio is shown in Figure \ref{fig:calib}(right) for three plausible critical values, $\bar{\psi} = 10^{-6}$, $10^{-4}$ or $10^{-3}$. The axis ratio evolution converges for low values of $\bar{\psi}$. 
We calibrate the lobe volume predicted by the RAiSE analytical model against that measured by the hydrodynamic simulation at $t = 0.33$\,Myr (i.e. peak in axis ratio of $A=7.6$), assuming $\bar{\psi} = 10^{-4}$, to derive an average value for the lobe filling factor of $\psi = 0.55$.

We model the rate at which the sound speed in the jet/lobe decreases after the commencement of lobe formation through the exponent on the `quadrature' sum in the lobe volume expression of Equation \ref{vol}. Following our method to fit the apparent half-opening angle of the jet, we evaluate the residual sum-of-squares between the lobe axis ratio of RAiSE and hydrodynamic simulation for each trial value of the exponent $\varsigma$ (cf. Equation \ref{RRS}). The RAiSE model prediction for the evolution of the lobe axis ratio for the best fit value of $\varsigma = 0.34$ (and the five parameters discussed above) is shown in Figure \ref{fig:calib}(right). Note that the low axis ratio ‘lobes’ predicted for $t < 0.33$\,Myr do not correspond to visible emission as $\Lambda < 0.1$, and thus a negligible fraction of the fluid is associated with either lobe expansion or emissivity. We consider $\Lambda = 0.1$ as the commencement of lobe formation in the discussion section.

\subsection{Dynamics of low-powered FR-I}
\label{sec:Dynamics of low-powered FR-I}

The calibrated RAiSE dynamical model (i.e. with parameters fitted in Section \ref{Sec:Free parameters in RAiSE}) reproduces the jet-head/lobe length expansion and axis ratio evolution of powerful \citeauthor{FR+1974} Type-II radio sources. However, it remains to be tested how much parameters can differ from those of the \citet{Yates+2022} hydrodynamic simulation whilst yielding accurate dynamical evolutionary histories. The computationally intensive nature of hydrodynamic simulations precludes a broad comparison across parameter space, so we therefore choose to simulate a single radio source with markedly different intrinsic parameters; a low-powered Fanaroff-Riley Type-I radio source with a `heavy', high-opening angle jet. 

The low-powered lobed FR-I radio source is simulated following the approach of \citet{Yates+2022}. Specifically, we use the relativistic hydrodynamics module of \texttt{PLUTO} version 4.3 to solve the fluid conservation equations on a three-dimensional Cartesian grid with second-order Runge–Kutta time-stepping, the HLLC Riemann solver, linear reconstruction, and the MINMOD limiter in the presence of shocks. The Taub-Mathews equation of state is used to relate fluid quantities \citep{Mathews+1971}. We assume the same spherically symmetric King profile as described in Section \ref{sec:Hydrodynamic Simulations}, and originally \citet{Yates+2021}. The jet injection region is defined as a sphere of radius 1\,kpc with the velocity set to $v_j = 0.01c$ (i.e. $\gamma_j = 1.00005$) along the two anti-parallel cones with half-opening angle $\theta_{j0} = 25$ degrees; the lobe axis ratio is independently set as $A = 2.82$ to match the simulation outputs. The jet power ($Q_{t\!\;\!o\!\;\!t} = 10^{36.3}$\,W, or $Q = 10^{36}$\,W) sets the density of the jet as $\rho_j(r) = 2Q/(v_j^3\Omega r^2)$, where $\Omega r^2$ is the cross-sectional area of the jet (cf. Section \ref{sec:Relativistic hydrodynamic equations}). The density of this low-powered jet at a radius of 1\,kpc is therefore a `heavy' $\rho_j = 1.32\times 10^{-22}\rm\,kg\,m^{-3}$ compared to $\bar{\rho}_j = 3.29\times 10^{-26}\rm\,kg\,m^{-3}$ for the high-powered jet of \citet{Yates+2022}.

\begin{figure*}
\begin{center}
\includegraphics[width=0.42\textwidth,trim={15 15 40 40},clip]{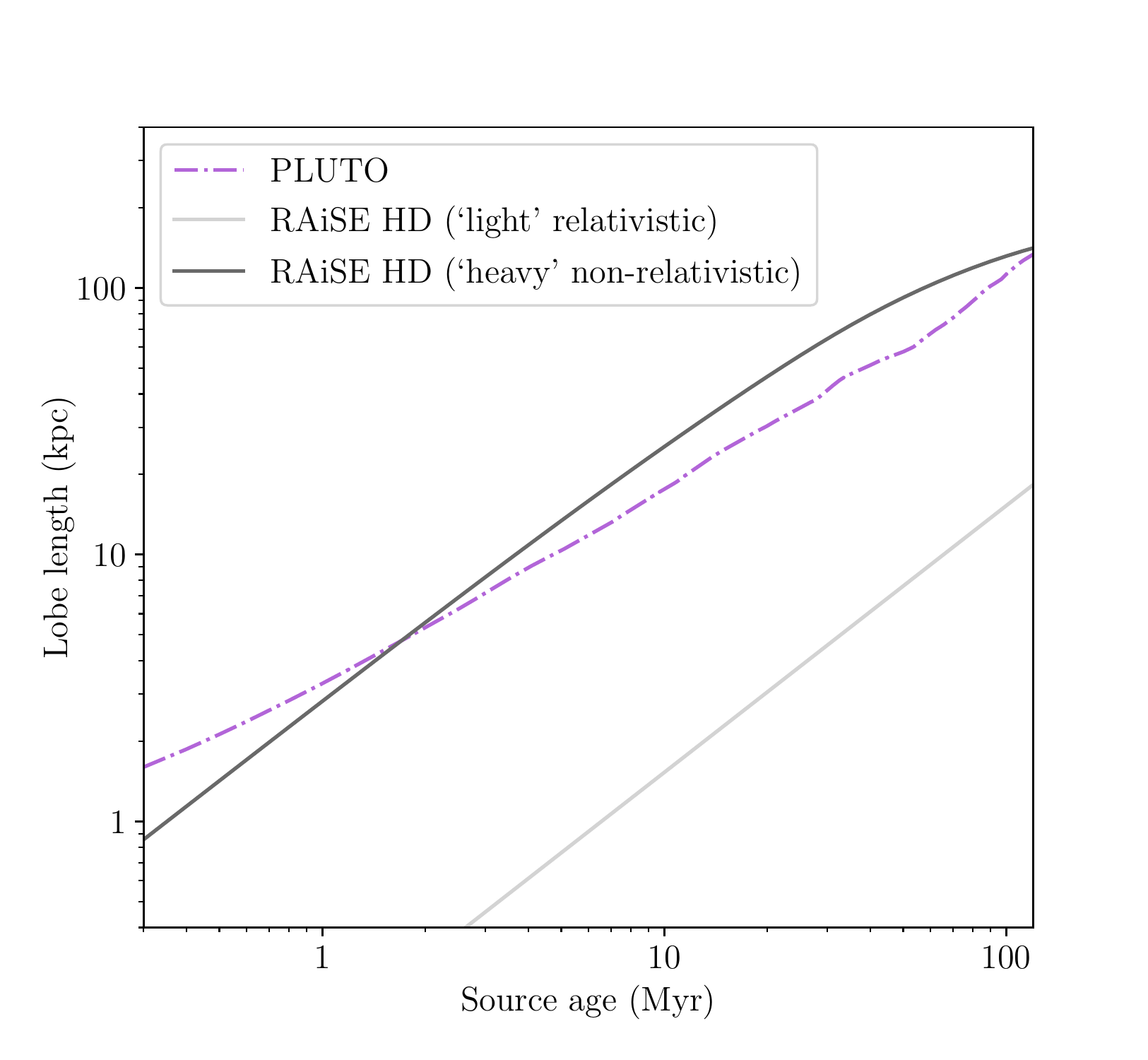} \quad\quad \includegraphics[width=0.42\textwidth,trim={15 15 40 40},clip]{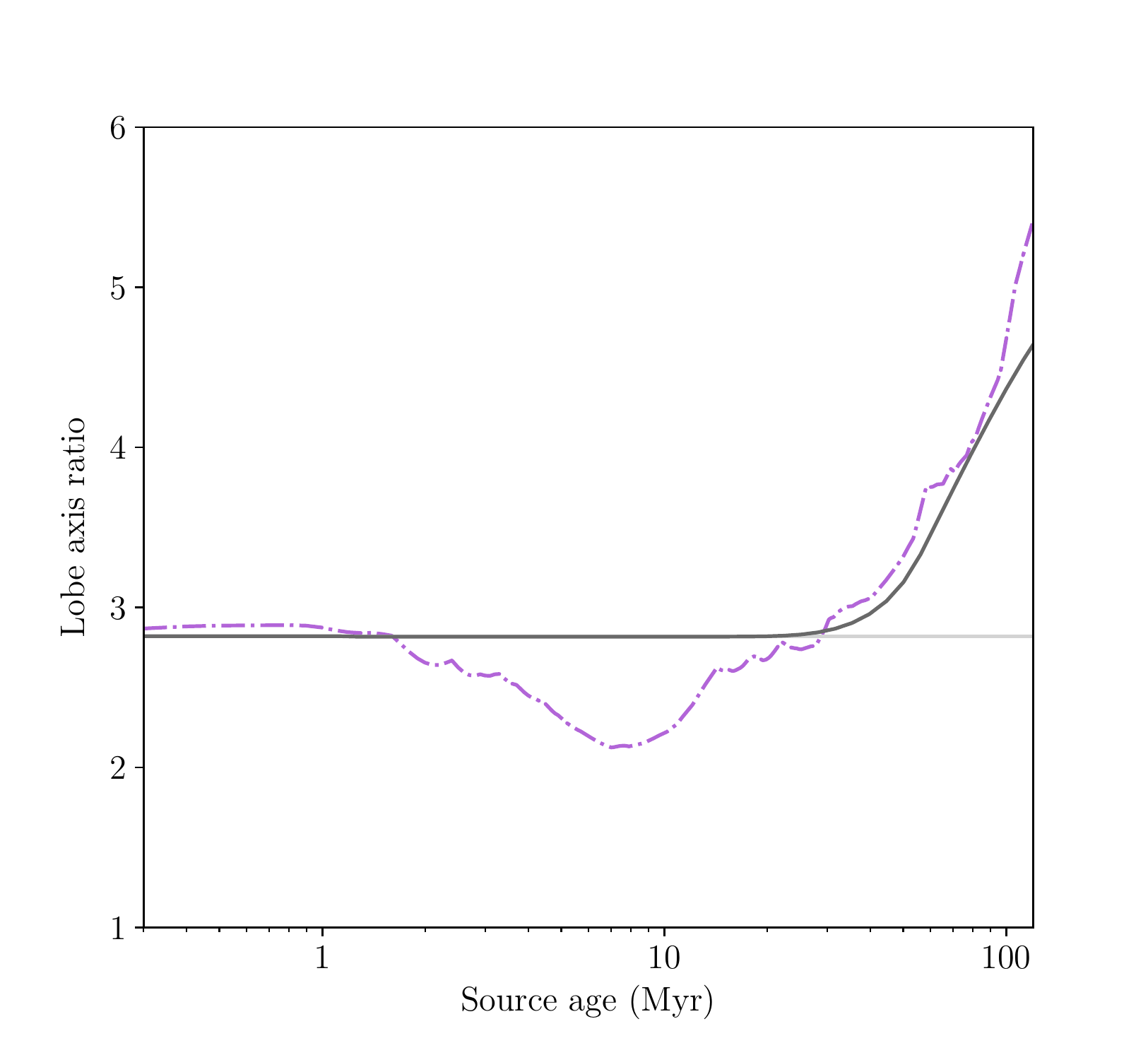}
\end{center}
\caption[]{Comparison of the calibrated RAiSE dynamical model predictions for a low-powered Fanaroff-Riley Type-I source compared to the (PLUTO) hydrodynamic simulation output for the corresponding parameters (see text); the RAiSE model is derived for either `light' relativistic (light grey) or `heavy' non-relativistic (dark grey) behaviour in the bulk flow of plasma along the jet. The jet-head/lobe length expansion (left) and axis ratio evolution (right) for both models are shown as a function of the source age. The axis ratio measured from the hydrodynamic simulation is unchanged for the range of lobe filling factors considered in Figure \ref{fig:calib}.}
\label{fig:fr1}
\end{figure*}

The evolutionary history of this low-powered radio source is derived using RAiSE for the same parameters as the hydrodynamic simulation, and the values constrained for the free model parameters in Section \ref{Sec:Free parameters in RAiSE} based on the high-powered radio source. However, the spine/sheath structure we assumed in Section \ref{sec:Jet spine and sheath} may not be relevant for `heavy' non-relativistic jets whose momentum flux is dominated by the rest-mass energy of the particles. We therefore consider both our previously discussed `light' relativistic fluid flow model for the jet (with density and velocity gradients across its cross-section), in addition to a `heavy' non-relativistic fluid flow model with an assumed constant density and velocity flow across the jet; i.e. $a_{*} = 1$. 
The jet-head/lobe length expansion and axis ratio evolution predicted by RAiSE and the (PLUTO) hydrodynamic simulation are compared in Figure \ref{fig:fr1}.
The close agreement for the `heavy' non-relativistic jet, especially the predicted time ($\sim$40\,Myr) and rate at which the axis ratio increases, confirms the validity of the RAiSE dynamical model over several orders of magnitude in intrinsic parameter space.

\subsection{Synthetic synchrotron emission images} 
\label{sec:Comparison of brightness images} 

The RAiSE dynamical model is now expected to produce comparable evolutionary histories to the hydrodynamic simulation of \citet{Yates+2022} as the six free model parameters were calibrated against their simulation of a jet expanding from the cluster centre. 
We now compare the surface brightness images produced by  the RAiSE model to the powerful FR-II hydrodynamic simulation\footnote{We do not compare surface brightness images for the low-powered FR-I as the seed Lagrangian particles used in this work are taken from an FR-II simulation.}; this not only validates the length and axis ratio (which we expect to match), but the pressure and magnetic field strength of the dynamical model, in addition to the spatial brightness distribution predicted by the Lagrangian particle-based emissivity model. Importantly, the Lagrangian particles used in this work are taken from 256 temporal snapshots of the \citet{Yates+2022} hydrodynamic simulation between a source age of 31.1 and 33.7\,Myr. As a result, we expect RAiSE brightness images at approximately $30$\,Myr to agree with the hydrodynamic simulation; however, the reproduction of surface brightness images at earlier times indicates the necessary fluid dynamics of the simulations are successfully captured in the RAiSE model. 

\begin{figure*}
\begin{center}
\includegraphics[width=0.9\textwidth,trim={75 80 20 90},clip]{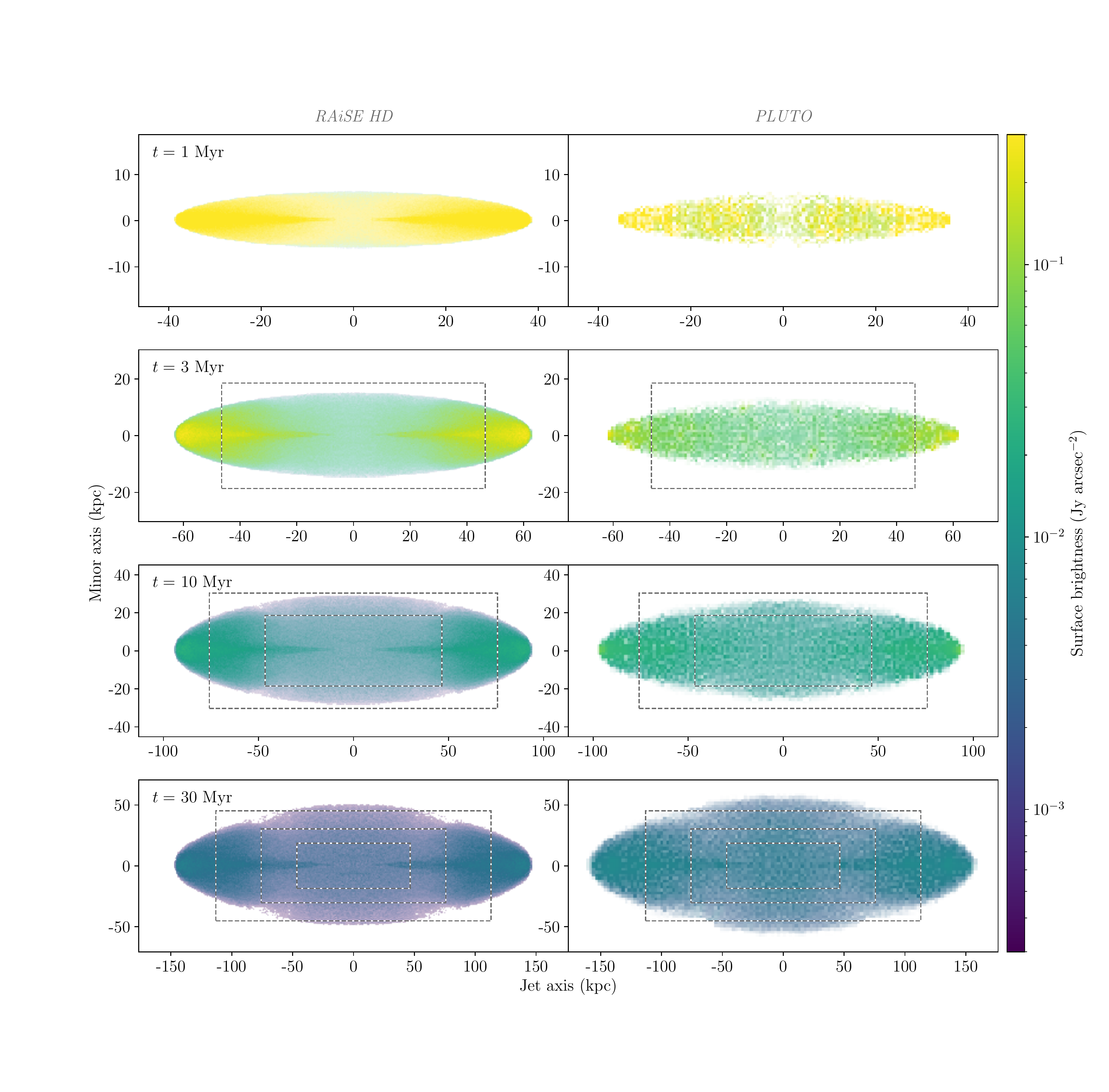}
\end{center}
\caption[]{Comparison of 1.4\,GHz (observer frame) radio-frequency surface brightness images produced by RAiSE (left) and the (PLUTO) hydrodynamic simulation (right) at redshift $z=0.05$. The resolution of RAiSE is greater than the hydrodynamic simulation as it combines Lagrangian particles from (an effective) 2048 temporal snapshots, resulting in more uniform coverage by the particles; the Lagrangian particles in RAiSE are taken from temporal snapshots at 31.1 to 33.7\,Myr.  The comparison is made at four ages throughout the source evolution: 1, 3, 10 and 30\,Myr. The dashed grey boxes shown in the bottom three panels show the plot boundaries for the earlier time steps. The radio sources are oriented in the plane of the sky.}
\label{fig:hydrocomp}
\end{figure*}

Radio-frequency surface brightness images from the RAiSE model and the (PLUTO) hydrodynamic simulation of \citet{Yates+2022} are compared in Figure \ref{fig:hydrocomp} at four times in their evolutionary history; 1\,Myr (top), 3\,Myr (second row), 10\,Myr (third row) and 30\,Myr (bottom). The same cluster environment and jet properties are assumed for both the model and simulation, so the images are directly comparable. The lobe lengths and axis ratios are consistent at the four time steps, as expected based on the success of their calibration, shown previously in Figure \ref{fig:calib}. Regardless, it is comforting that the shapes of the visible lobes are consistent, not just their lengths and widths.

The radio-frequency emissivity in Figure \ref{fig:hydrocomp} is calculated at 1.4\,GHz (observer frame) for a redshift of $z=0.05$, with the surface brightness scaled assuming a 1\,arcsec$^2$ beam. The spatial brightness distribution predicted by RAiSE at 30\,Myr (bottom-left) is unsurprisingly consistent with that of the hydrodynamic simulation (bottom-right), given the seed Lagrangian particles are taken from close to this time step. The resolution of the RAiSE image is significantly improved over the original simulation as it combines particles from 256 proximate temporal snapshots and randomly perturbs each into eight different orientations (i.e. 2048 effective temporal snapshots). The difference in resolution is most pronounced at earlier times as fewer hydrodynamic simulation particles have been injected into the lobe; the noise in Figure \ref{fig:hydrocomp}(top-right) results from very few (if any) particles per pixel. By contrast, the RAiSE model consistently uses a full set of 115 million particles at all time steps (assuming the highest resolution is specified). Regardless of the different resolutions, examining Figure \ref{fig:hydrocomp}, it is apparent that RAiSE produces very comparable spatial brightness distributions to the hydrodynamic simulation for the 1, 3 and 10\,Myr time steps.

\section{Jet contribution to evolutionary history} 
\label{sec:Jet contribution to evolutionary history} 

The importance of the jet expansion phase, added to the RAiSE dynamical model in this work, is investigated by comparing to a simplified model using solely the differential equations describing lobe dynamics. This simplified dynamical model is identical to that of the earlier RAiSE X model \citep{Turner+2020a}, however the particle based emissivity calculation developed in this work is retained to enable a direct comparison of their relative luminosities. 

\subsection{Radio source evolutionary tracks}

The RAiSE dynamical model is used to predict the evolutionary history of a radio source with properties matching the hydrodynamic simulation of \citet{Yates+2022}; this is the base model considered in this section. The properties of this radio source are detailed in Section \ref{sec:Hydrodynamic Simulations}. This radio source is further modelled for the same intrinsic parameters (i.e. jet power, late-time axis ratio, cluster environment, etc.) but without including the jet-dominated expansion phase in the RAiSE model.

\subsubsection{Lobe dynamics}
\label{sec:Lobe dynamics}

\begin{figure*}
\begin{center}
\includegraphics[width=0.99\textwidth,trim={85 10 105 40},clip]{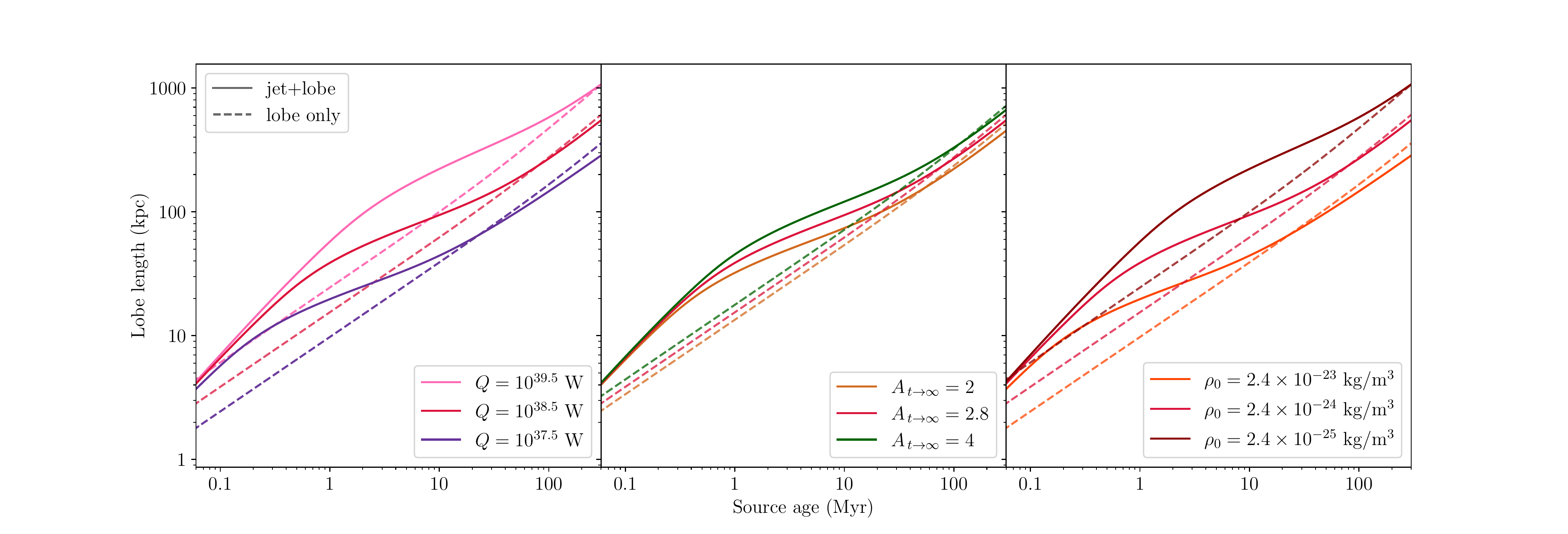}
\end{center}
\caption[]{Modelled expansion of the jet-head/lobe length (single lobe) as a function of the source age using the RAiSE model either including (solid lines) or excluding (dashed lines) a jet-dominated expansion phase. The evolution of radio source shown in red (base model) assumes the same intrinsic parameters as the cluster-centred hydrodynamic simulation of \citet{Yates+2022}. The jet power, late-time lobe axis ratio and core gas density are modified (relative to this base model) in the left, centre and right panels respectively.}
\label{fig:jetcomp_length}
\end{figure*}

The expansion of the jet-head/lobe length for models including (`jet+lobe') or excluding (`lobe only') the jet-dominated expansion phase are compared in Figure \ref{fig:jetcomp_length} (solid red and dashed red lines respectively).
The inclusion of the jet-dominated expansion phase results in substantially higher advance speeds prior to the commencement of lobe formation at $t = 0.33$\,Myr (see Section \ref{sec:Lobe filling factor and sound speed}). As a result, this `jet+lobe' model predicts a jet-head/lobe length of 21.7\,kpc at lobe formation compared to 8.9\,kpc (at the same age) for the `lobe only' model. 
However, the forward expansion slows quickly upon the commencement of lobe formation in the `jet+lobe' model whilst the strong-shock supersonic expansion in the `lobe only' model continues to slow gently \citep[cf.][their Equation 4]{KA+1997}. The lobe length evolution predicted by both models converges at late-times in the lobe-dominated expansion phase at $t \geqslant 60$\,Myr; the initial differences in the evolutionary histories are inconsequential for the oldest radio source populations. 

\begin{figure*}
\begin{center}
\includegraphics[width=0.99\textwidth,trim={85 10 105 40},clip]{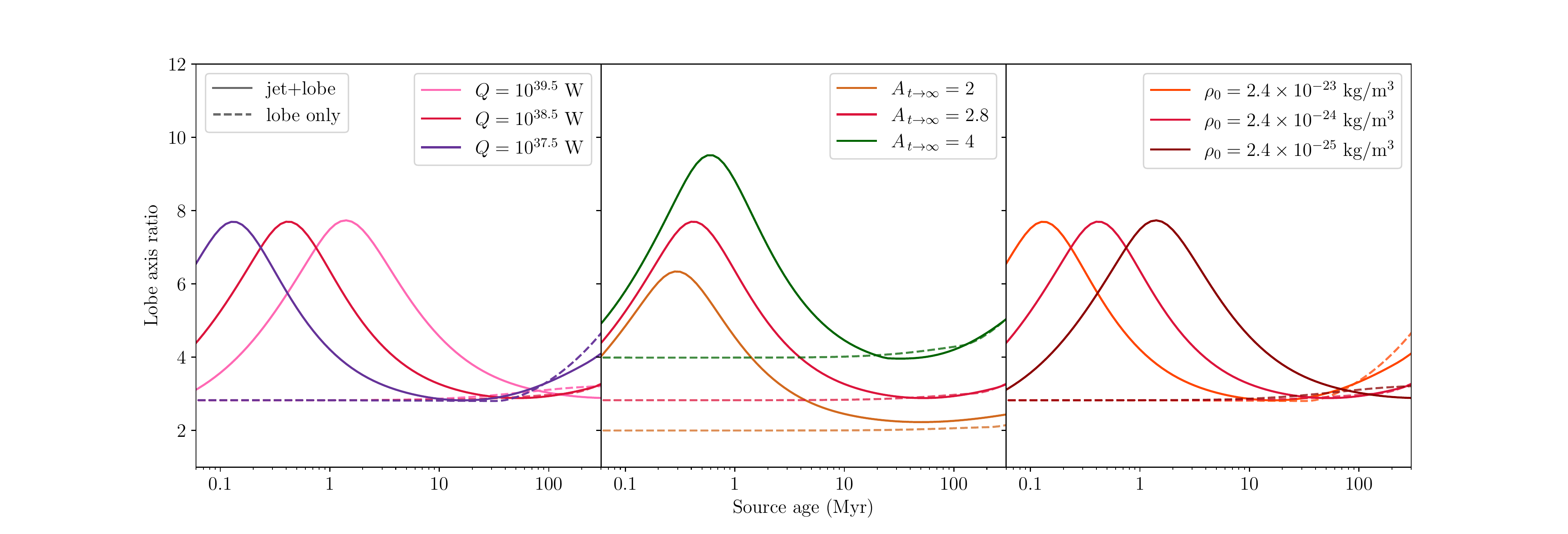}
\end{center}
\caption[]{Same as Figure \ref{fig:jetcomp_length} but for the evolution in the lobe axis ratio as a function of the source age.}
\label{fig:jetcomp_axis}
\end{figure*}

\begin{figure*}
\begin{center}
\includegraphics[width=0.99\textwidth,trim={85 10 105 40},clip]{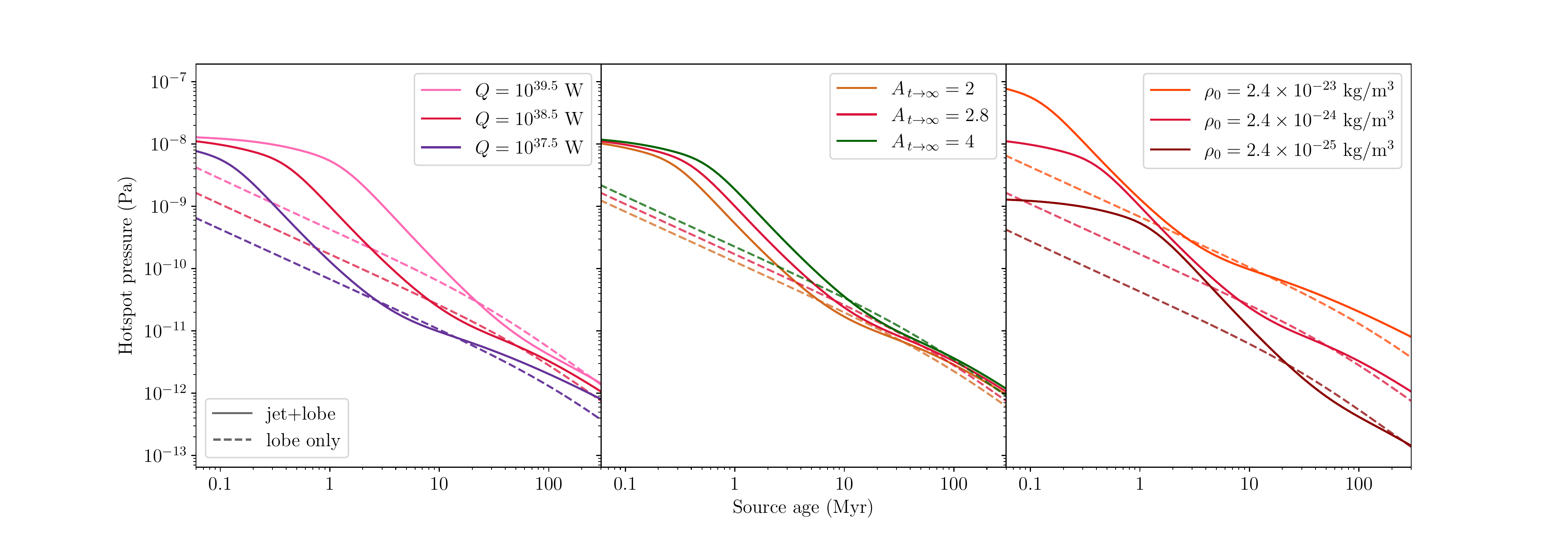}
\end{center}
\caption[]{Same as Figure \ref{fig:jetcomp_length} but for the evolution in the jet-head/hotspot pressure (ram and thermal components) as a function of the source age.}
\label{fig:jetcomp_pressure}
\end{figure*}

The `jet+lobe' model shows considerable variation in the lobe axis ratio throughout the radio source evolutionary history (solid red line in Figure \ref{fig:jetcomp_axis}); the lobe is initially quite narrow at the commencement of lobe formation (i.e. peak in axis ratio of $A=7.6$ at $t = 0.33$\,Myr) before widening to its late-time value of $A=2.83$ at approximately $30$\,Myr. By contrast, the `lobe only' model (dashed red line) maintains a constant axis ratio throughout the strong-shock supersonic expansion phase. Both models predict a slight narrowing of the lobe (and shocked shell) after 100\,Myr due to a combination of weakly supersonic expansion and a steepening ambient gas density profile \citep[see][]{Turner+2015}; we note that the RAiSE model does not consider buoyancy which would ultimately result in the lobes being separated (or `pinched') from the AGN core. Meanwhile, the jet-head/hotspot pressure (ram and thermal components; Figure \ref{fig:jetcomp_pressure}) is predicted to take an approximately constant value in the jet-dominated expansion phase, before rapidly decreasing upon the commencement of lobe formation. The `lobe only' model predicts a significantly lower pressure at these times, but both models (approximately) converge to the same pressure at $\geqslant10$\,Myr. {Importantly, the lobe axis ratios presented in Figure \ref{fig:jetcomp_axis} do not consider survey surface brightness sensitivity; the fainter emission away from the jet axis, a result of radiative losses in the aged electron population, would not be detected in the oldest sources leading to much higher axis ratios in the observed population \citep{Turner+2018a}.} 

We additionally investigate the magnitude of changes to the lobe formation length-scale for a range of jet powers, late-time axis ratios, and core gas densities (albeit with the same profile shape assumed by \citealt{Yates+2022}). Higher jet kinetic powers ($Q = 10^{39.5}$\,W) result in the lobe formation occurring for an older source age ($\sim$1.05\,Myr) and greater galactocentric radius (75\,kpc), and conversely for lower jet powers ($Q = 10^{37.5}$\,W), lobe formation occurs earlier ($\sim$0.1\,Myr) closer to the central nucleus (7.5\,kpc); see Figures \ref{fig:jetcomp_length}(left) and \ref{fig:jetcomp_axis}(left). This strong relationship is expected as the jet density increases linearly with increasing kinetic power (Equation \ref{jetdensity}), thus maintaining the stability of the flow for greater distances (see Equation \ref{bigL} for $\mathcal{L}$). 
Meanwhile, the apparent half-opening angle of the jet, modified through the late-time axis ratio of the lobe (Equation \ref{axisopening}), results in only minor changes to the lobe formation length-scale; however, narrower jets ($A=4$) are associated with slightly later lobe formation than wider jets ($A=2$), as shown in Figure \ref{fig:jetcomp_length}(centre). The lobe axis ratios unsurprisingly vary for different jet opening angles but converge to the specified late-time value; see Figure \ref{fig:jetcomp_axis}(centre).
Finally, varying the core gas density has the opposite result on the lobe formation length-scale as the jet power; see Figure \ref{fig:jetcomp_length}(right). This occurs as the metric describing the commencement of lobe formation, $\mathcal{L}$ (Equation \ref{bigL}), is linearly proportional to jet power but inversely proportional to the ambient gas density. Importantly, the pressure of the jet-head/hotspot (ram and thermal components) scales linearly with the core gas density for an otherwise identical profile shape; see Figure \ref{fig:jetcomp_pressure}(right).

\subsubsection{{Radio luminosity}}

\begin{figure*}
\begin{center}
\includegraphics[width=0.99\textwidth,trim={85 10 105 40},clip]{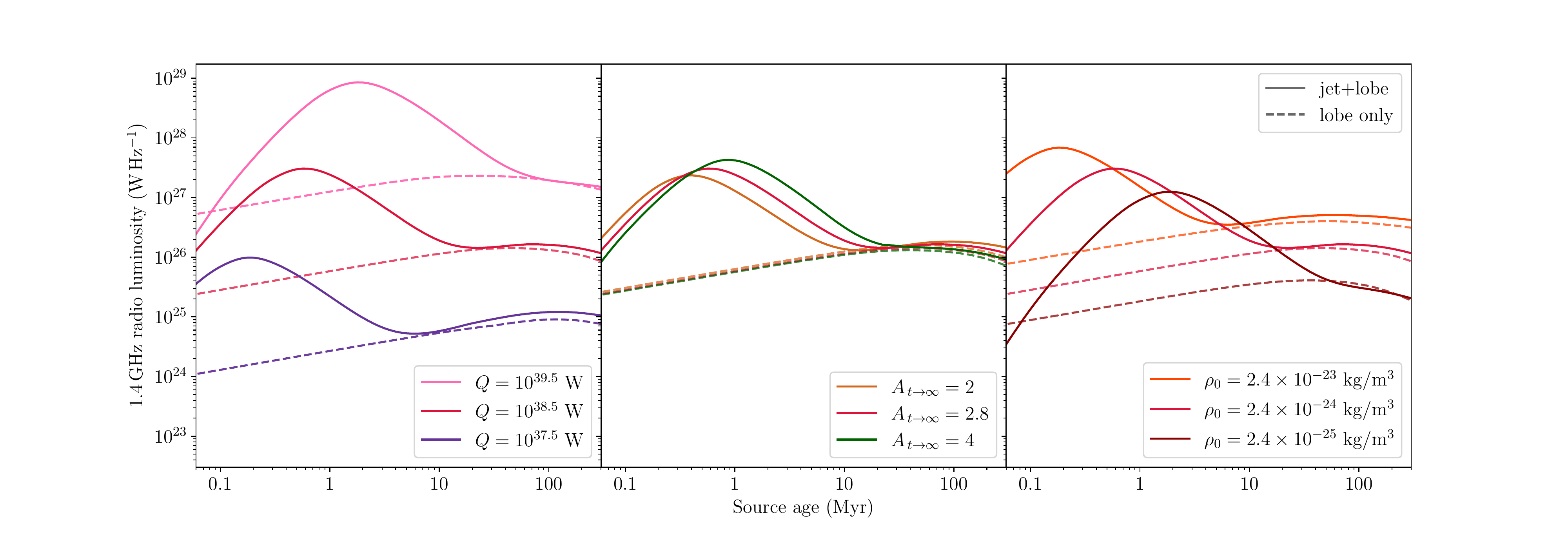}
\end{center}
\caption[]{Modelled evolution of the integrated 1.4\,GHz (observer frame) radio-frequency luminosity from a single jet/lobe as a function of the source age using the RAiSE model. These radio sources are modelled at redshift $z=0.05$; see \citet{Yates+2022} for a discussion of the simulated luminosity tracks at other frequencies and higher redshifts. See Figure \ref{fig:jetcomp_length} for complete description.}
\label{fig:jetcomp_radio}
\end{figure*}

We compare the evolution of the integrated radio-frequency luminosity for the `jet+lobe' and `lobe only' models in Figure \ref{fig:jetcomp_radio} (solid red and dashed red lines respectively). The radio-frequency emissivity is calculated at 1.4\,GHz (observer frame) for a redshift of $z = 0.05$; we refer the interested reader to \citet[][their Figure 8]{Yates+2022} for a discussion of simulated luminosity tracks at other frequencies and higher redshifts. The luminosity of the radio source rises rapidly in the jet-dominated expansion phase, quickly surpassing the comparatively constant brightness predicted for the `lobe only' model \citep[cf.][]{KDA+1997,Turner+2015}. The radio-frequency emissivity peaks at 0.6\,Myr shortly after the commencement of lobe formation (at $t=0.33$\,Myr). Beyond this peak, the emissivity reduces due to two factors: (1) the thermal pressure continues to reduce at a similar rate, i.e. $p \propto t^{(-4-\beta)/(5-\beta)}$; but (2) the forward expansion of the jet-head/lobe is greatly reduced in the lobe-dominated expansion phase, leading to a slower volume growth rate once the lobe is partially formed (i.e. for $t > 0.6$\,Myr). This rapid increase, and subsequent fall, in luminosity is consistent with the earlier hydrodynamic simulations of \citet{HK+2013} that assume a simple scaling between pressure and radio emissivity (their Figure 13). The RAiSE `jet+lobe' luminosity tracks converge to the `lobe only' prediction at late-times ($t > 10$\,Myr); both models show a continued increase in emissivity for a short period before succumbing to increased radiative losses beyond 100\,Myr. 

{We investigate} changes to the luminosity tracks for a range of jet powers, late-time axis ratios, and core gas densities. The commencement of lobe formation (which approximately corresponds to the peak luminosity), shifts with these intrinsic parameters as discussed in Section \ref{sec:Lobe dynamics}. Further, the {synchrotron emissivity scales} linearly with both the jet power and core gas density; {see Figure \ref{fig:jetcomp_radio}. }Higher jet powers are expected to increase the integrated luminosity of the source because, as discussed in Section \ref{sec:Lobe dynamics}, increasing the jet power raises the lobe volume (Figures \ref{fig:jetcomp_length} and \ref{fig:jetcomp_axis}) whilst maintaining the thermal pressure (and thus magnetic field strength). Meanwhile, greater gas densities directly increase the thermal pressure in the lobe, resulting in {increased brightness. These} findings are qualitatively identical to well-established lobe emissivity models {\citep[e.g.][]{KDA+1997, Turner+2018a, Hardcastle+2018, Turner+2020a}.}

\subsubsection{Size--luminosity tracks in {local galaxy} clusters}

\begin{figure*}
\begin{center}
\includegraphics[width=0.99\textwidth,trim={85 10 105 40},clip]{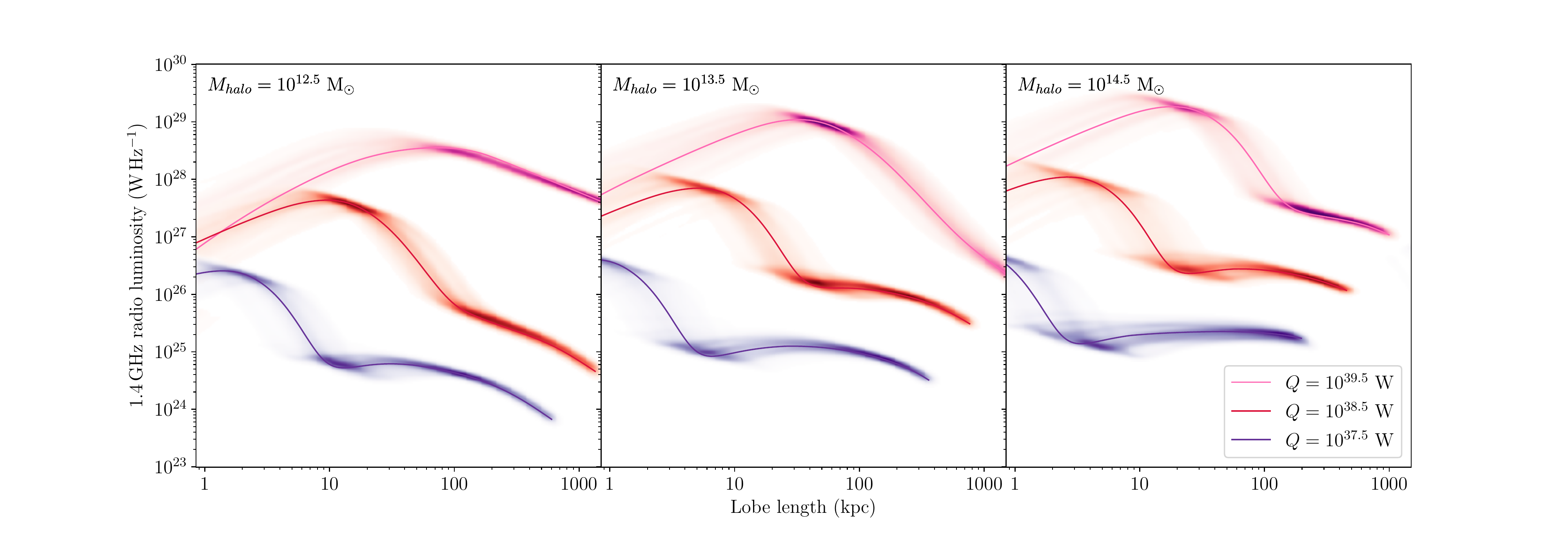}
\end{center}
\caption[]{Modelled evolution of the integrated 1.4\,GHz (observer frame) radio-frequency luminosity from a single jet/lobe as a function of the jet-head/lobe length using the RAiSE model. Size--luminosity tracks are modelled for three jet powers and with gas density profiles based on {the variation seen in X-ray observations of galaxy clusters }by \citet{Vikhlinin+2006}. The gas density profiles assume halo masses representative of either a small group (left; $M_{h\!\;\!a\!\;\!l\!\;\!o} = 10^{12.5}$\,M$_\odot$), a large group (centre; $M_{h\!\;\!a\!\;\!l\!\;\!o} = 10^{13.5}$\,M$_\odot$) or a cluster (right; $M_{h\!\;\!a\!\;\!l\!\;\!o} = 10^{14.5}$\,M$_\odot$). See Figure \ref{fig:jetcomp_radio} for complete description.}
\label{fig:jet_env}
\end{figure*}

{The analysis performed so far has assumed a single gas density profile for the ambient medium. We now investigate the sensitivity of our findings by comparing the stability of predicted size--luminosity tracks to {the diversity of gas density profiles obtained from X-ray observations of local galaxy clusters.} Following \citet[][their Equation 13]{Turner+2015}, we describe the ambient gas density profile as simplified form of the double-$\beta$ profile derived by \citet{Vikhlinin+2006} with the core density and virial radius scaled for a given halo mass based on the outputs of semi-analytic galaxy evolution models \citep[SAGE][]{Croton+2006}. Meanwhile, we consider variation in the profile shape by pseudo-randomly sampling values from the measured probability density functions for each of the six parameters describing the double-$\beta$ profile. We generate a set of 100 profile shapes for each tested halo mass, and model the jet/lobe evolution for these sets of environments for a range of jet powers.}

{We now investigate the variation in the resulting size--luminosity tracks due to the different profile shapes, for an AGN with a given jet power and halo mass environment. The evolutionary tracks are used to calculate the relative likelihood of observing the radio lobe at a given location in size--luminosity parameter space; i.e. we assign the size--luminosity pair for each log-space time step to a grid of width $0.05$\,dex in both dimensions. The corresponding likelihood distributions in size--luminosity parameter space are shown in Figure \ref{fig:jet_env} for a range of jet powers; the gas density profiles assume halo masses representative of either a small group (left; $M_{h\!\;\!a\!\;\!l\!\;\!o} = 10^{12.5}$\,M$_\odot$), a large group (centre; $M_{h\!\;\!a\!\;\!l\!\;\!o} = 10^{13.5}$\,M$_\odot$) or a cluster (right; $M_{h\!\;\!a\!\;\!l\!\;\!o} = 10^{14.5}$\,M$_\odot$).}

{The late-time evolution of the lobe has remarkably consistent size--luminosity tracks between the breadth of modelled profile shapes for a given jet power and halo mass. This is somewhat expected given the virial radius, comparable to the size of the late-time lobes, is strongly correlated with halo mass. By contrast, the lobe formation length-scale is highly sensitive to the shape of the profile in the cluster core, most notably for lower jet powers ($Q = 10^{37.5}$\,W; purple) and in the high-mass cluster environment ($M_{h\!\;\!a\!\;\!l\!\;\!o} = 10^{14.5}$\,M$_\odot$; right). The uncertainty in the lobe length is, however, at most a factor of two and therefore no more significant than other parameters such as the axis ratio to accurately describe the size--luminosity evolution. Consequently, the expected variability in gas density profiles, for a given halo mass environment, adds minimal additional uncertainty to the jet power and source age estimates obtained in parameter inversions \citep[cf.][]{Turner+2015,Turner+2018b}. }

\subsection{Surface brightness images}

\begin{figure*}
\begin{center}
\includegraphics[width=0.9\textwidth,trim={75 80 20 90},clip]{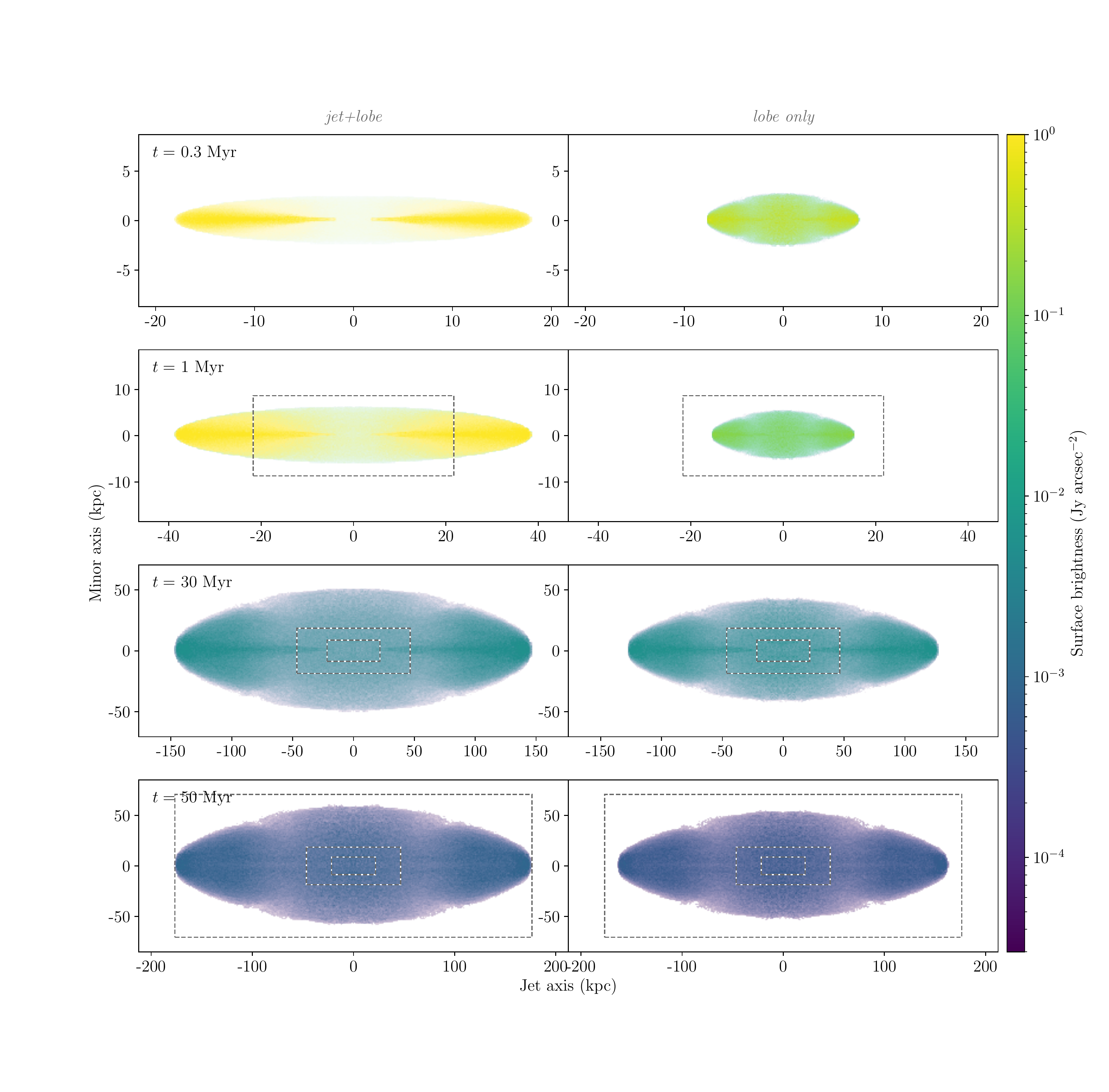}
\end{center}
\caption[]{Comparison of 1.4\,GHz (observer frame) surface brightness images produced by RAiSE at redshift $z=0.05$ either including (left) or excluding (right) a jet-dominated expansion phase; the active nucleus switches off after 30\,Myr in both models. The comparison is made at four ages throughout the source evolution: 0.3, 1, 30 and 50\,Myr. See Figure \ref{fig:hydrocomp} for complete description.}
\label{fig:jetcomp}
\end{figure*}

The spatial distribution of the {synchrotron emissivity} is, in practice, more important than the integrated luminosity when considering the surface brightness sensitivity of radio surveys. We compare modelled radio-frequency images for the `jet+lobe' and `lobe only' models at four key stages in the radio source evolutionary history: jet-dominated expansion, lobe formation, lobe-dominated expansion, and the cessation of jet activity (i.e. remnant phase). For both models, we consider a $Q = 10^{38.5}$\,W jet emanating from the centre of the \citet{Yates+2021} cluster profile for an active age of $t_{o\!\;\!n} = 30$\,Myr. We generate surface brightness images assuming 1.4\,GHz (observer frame) emission for a radio source at redshift $z=0.05$. The radio-frequency images are shown in Figure \ref{fig:jetcomp} for the `jet+lobe' and `lobe only' models at 0.3, 1, 30 and 50\,Myr, probing each of the four previously identified stages in the radio source evolutionary history.

The spatial brightness distribution in Figure \ref{fig:jetcomp}(top-left) evidently arises from towards the end of the jet-dominated expansion phase, with very strong synchrotron emission arising from the jets and jet-head, but negligible emission from the surrounding `lobe'. The assumption of a two-phase fluid (see Section \ref{sec:Two phase fluid}) to model the transition from a jet- to lobe- dominated flow will always associate some level of emission with the lobe component. However, for even younger sources (where $\Lambda \ll 0.1$), this modelled emission is too faint to be detected by any radio survey. By contrast, the `lobe only' model predicts significantly higher surface brightness across the ellipsoidal radio source and a lobe length that is approximately half that of the jet in the `jet+lobe' model (cf. Figure \ref{fig:jetcomp_length}). The surface brightness images at 1\,Myr lead to similar findings, but we note that backflow of synchrotron-emitting plasma has now filled the narrow, partially formed lobe; see Figure \ref{fig:jetcomp}(second row-left).

The third key stage of the radio source evolutionary history we investigate is the late-time expansion of the lobe; see Figure \ref{fig:jetcomp}(third row). The lobe of the `jet+lobe' model is fully formed at this time with approximately uniform surface brightness, though with enhanced emission around the jets and hotspots. The axis ratio of the lobe has converged to the specified late-time value of $A=2.83$. The `lobe only' predicts a very comparable integrated luminosity (cf. Figure \ref{fig:jetcomp_radio}), but marginally smaller lobe length (cf. Figure \ref{fig:jetcomp_length}), leading to a slightly higher surface brightness. Finally, examining the remnant phase of the source evolution in Figure \ref{fig:jetcomp}(bottom), we find very comparable lobe lengths, luminosities and surface brightnesses. Notably, there is no longer any emission associated with the jet (Equation \ref{remjet}), whilst the hotspot region has faded more rapidly than the rest of the lobe plasma, leading to a spatially uniform brightness distribution across the lobe.

\section{CONCLUSIONS}
\label{sec:CONCLUSIONS}

We have presented a new analytical radio galaxy model which includes an early jet-dominated growth phase, followed by the formation of lobes, and subsequent standard lobe-dominated expansion phase; this model is implemented within the established \emph{Radio AGN in Semi-Analytic Environments} (RAiSE) framework.
Jet-phase parameters describing the behaviour of the relativistic fluid flow are calibrated using integrated quantities from hydrodynamic simulations (i.e. jet-head/lobe length and axis ratio).
We find that {our new RAiSE model} reproduces the evolution of dynamical quantities in hydrodynamic simulations of both Fanaroff-Riley
Type-I and -II radio lobes. 
However, omitting the jet-dominated expansion phase, as for existing analytical models, results in an underprediction of the source size and luminosity at a given jet power; or conversely, an overprediction of the jet kinetic power inferred from the size and radio luminosity in parameter inversions \citep[e.g.][]{Shabala+2008,Turner+2015,Hardcastle+2019a}. These effects are most dominant for young ($\lesssim 10$\,Myr) sources, with the exact transition from jet- to lobe-dominated expansion dependent on jet and environment parameters. 

We produce {synthetic synchrotron surface brightness} images of Fanaroff-Riley Type-II radio sources by modifying the dynamics of Lagrangian tracer particles taken from an existing hydrodynamic simulation; these particles trace the magnetic field and shock-acceleration histories throughout the lobe plasma. We show that a single set of particles, when adapted to the dynamics of the RAiSE analytical model, is sufficient to accurately reproduce the spatial distribution of dynamical quantities and synchrotron emissivities in hydrodynamic simulations. \citet{Quici+2022} demonstrated that surface brightness images at just two radio frequencies, as expected from all sky surveys, provide orthogonal constraints on at least four intrinsic source properties in parameter inversions: kinetic jet power, active age (`on time'), remnant age (`off time'), and magnetic field strength. The computational simplicity of RAiSE, and comparable accuracy of model predictions to hydrodynamic simulations, makes it a promising tool for interpretation of AGN populations in large radio surveys. 

The RAiSE code is publicly available on GitHub and PyPI (see \hyperref[sec:DATA AVAILABILITY]{Data Availability}).

\section*{ACKNOWLEDGEMENTS}

We thank Seth N. Kaehne for his contribution to the online documentation provided with the code as part of the University of Tasmania Dean's Summer Research program{, and an anonymous referee for their constructive comments that have improved the manuscript.}

\section*{\phantomsection
DATA AVAILABILITY}
\label{sec:DATA AVAILABILITY}

The authors confirm that the data supporting the findings of this study are available within the article and the relevant code is is publicly available on GitHub (\href{https://github.com/rossjturner/RAiSEHD}{github.com/rossjturner/RAiSEHD}) and PyPI (\href{https://pypi.org/project/RAiSEHD}{pypi.org/project/RAiSEHD}).



\label{lastpage}
\end{document}